\documentclass{aa}  

\usepackage{graphicx}
\usepackage{txfonts}
\usepackage{soul}  

\usepackage{hyperref}
\hypersetup{
    colorlinks=true,
    linkcolor=blue,
    filecolor=magenta,      
    urlcolor=cyan,
    citecolor=blue,
    pdfpagemode=UseThumbs, 
    }
\usepackage{makecell}
\usepackage{float}
\usepackage{natbib}
\usepackage{xcolor}
\usepackage[normalem]{ulem}
\usepackage{placeins} 
\usepackage{stfloats} 
\bibpunct{(}{)}{;}{a}{}{,} 
\usepackage{afterpage} 

\defcitealias{2023paperI}{Paper I}

\newcommand{\obd}{18OB}
\newcommand{\obc}{4OB}
\newcommand{\grey}{grey}

\begin{document} 
   \title{Hydrodynamic simulations of cool stellar atmospheres with \texttt{MANCHA}} 

   \author{A. Perdomo García
          \inst{1}\fnmsep\inst{2}
          \and
          N. Vitas\inst{1}\fnmsep\inst{2}
          \and 
          E. Khomenko
          \inst{1}\fnmsep\inst{2}
          \and 
          M. Collados
          \inst{1}\fnmsep\inst{2}
          }

   \institute{Instituto de Astrofísica de Canarias,
              38200 La Laguna, Tenerife, Spain 
              \\
              \email{andrperdomo@gmail.com}
         \and
             Departamento de Astrofísica de la Universidad de La Laguna,
             38200 La Laguna, Tenerife, Spain
             }

   \date{Received Month day, year; accepted Month day, year}


  \abstract
   {Three-dimensional time-dependent simulations of stellar atmospheres are essential to study the surface of stars other than the Sun. These simulations require the opacity binning method to reduce the computational cost of solving the radiative transfer equation down to viable limits. 
   The method depends on a series of free parameters, among which the location and number of bins are key to set the accuracy of the resulting opacity. }
   {Our aim is to test how different binning strategies previously studied in one-dimensional models perform in three-dimensional radiative hydrodynamic simulations of stellar atmospheres. }
   {Realistic box-in-a-star simulations of the near-surface convection and photosphere of three spectral types (G2V, K0V, and M2V) were run with the \texttt{MANCHA} code with grey opacity. 
   After reaching the stationary state, one snapshot of each of the three stellar simulations was used to compute the radiative energy exchange rate with grey opacity, opacity binned in four $\tau$-bins, and opacity binned in 18 $\{ \tau, \lambda \}$-bins. These rates were compared with the ones computed with opacity distribution functions. 
   Then, stellar simulations were run with grey, four-bin, and 18-bin opacities to see the impact of the opacity setup on the mean stratification of the temperature and its gradient after time evolution.
   }
   {The simulations of main sequence cool stars with the \texttt{MANCHA} code are consistent with those in the literature. 
   For the three stars, the radiative energy exchange rates computed with 18 bins are remarkably close to the ones computed with the opacity distribution functions. The rates computed with four bins are similar to the rates computed with 18 bins, and present a significant improvement with respect to the rates computed with the Rosseland opacity, especially above the stellar surface. The Rosseland mean can reproduce the proper rates in sub-surface layers, but produces large errors for the atmospheric layers of the G2V and K0V stars. In the case of the M2V star, the Rosseland mean fails even in sub-surface layers, owing to the importance of the contribution from molecular lines in the opacity, underestimated by the harmonic mean. Similar conclusions are reached studying the mean stratification of the temperature and its gradient after time evolution. }
   {}

   \keywords{Hydrodynamics --
             Plasmas --
             Stars: atmospheres --
             Convection --
             Opacity
               }

   \maketitle
%


\section{Introduction} \label{sec:introduction}

    Stars can only be accessed through their radiation, for which stellar models are key to understand their underlying physics. 
    One-dimensional (1D) models of stellar atmospheres are commonly constructed imposing hydrostatic equilibrium, plane-parallel stratification, and mixing length theory to account for the effect of convection in the energy transport (with the need of ad hoc parameters). Regardless of their success reproducing many observables (see e.g. \citealp{2003gustafsson_IAU} and references therein), the models are limited by their 1D nature, and any three-dimensional (3D) phenomenon that happens in the stellar atmospheres, needs to be approximately treated (e.g. line broadening) or is impossible to reproduce (e.g. convective blueshifts). Moreover, although the models are capable of reproducing the spectrum of an atmosphere, they do not represent its real mean stratification \citep{2011_Uitenbroek_1Dfails}.

    On the contrary, 3D magneto-hydrodynamic (MHD) simulations of stellar atmospheres naturally reproduce realistic convective motions \citep{1998SteinNordlund} and replicate or even improve the predictions of 1D models \citep{2010gustafsson_realistic}, such as line asymmetries \citep{1990Dravins_lineSimulations, 1990Dravins_lineSimulations_II}, solar \citep{A09, 2011Caffau, 2021_Bergemann_SolarOxygen, 2021Amarsi, 2021Asplund} and stellar \citep{2019Amarsi} abundances, or centre-to-limb variation (CLV, \citealp{2013_Pereira_solar_atms}). 
    These 3D simulations are confirmed to represent the solar atmosphere realistically through several comparisons against the observations \citep{2009_solarconvection_Nordlund}. Thus, they allow us to study in detail the atmosphere of distant stars that are unreachable directly through observations.
    
    While the simulations require large data cubes and computational power, 1D models are less costly and can afford for a detailed treatment of radiation, including millions of transitions and wavelengths in the solution of the radiative transfer equation (RTE). In the 3D simulations such a detailed radiative transfer (RT) is impossible with the current computing power, and different strategies are designed to reduce the computational expense, maintaining a satisfactory representation of the wavelength dependence (so the bolometric radiative outputs are accurate enough). First, the opacity is precomputed in lookup tables (dependent on two thermodynamic variables, the chemical composition and turbulent velocity) under the assumption of local thermodynamic equilibrium (LTE). Second, a further approximation can be obtained by combining the opacity into a few representative values that reproduce the bolometric radiative outputs accurately enough. Some examples of these kinds of approximations are the opacity sampling \citep{1976Sneden_OS}, opacity distribution functions (ODFs, \citealp{1951labs_ODF}, \citealp{ODF1966}), and opacity binning (OB, \citealp{norlund1982}). Only the OB enables sufficient speedup in 3D simulations. 
    Among the choices to make in the OB method, the number and location of the bins highly determines the accuracy of the approximation. Several authors have experimented with a different number and distribution of the bins, 
    using bins grouping the opacities in optical depth $\tau$ (\citealp{norlund1982}, \citealp{voegler_thesis}, \citealp{beeck_thesis}), or including additional splitting of the opacities in wavelength $\lambda$ (\citealp{ludwig_thesis}, \citealp{2013_Pereira_solar_atms}, \citealp{magic_thesis}, \citealp{2018Collet}, \citealp{2023Zhou}). 
    As studied in \citet{2004voegler_nongrey}, for applications focussed at magnetoconvective processes close to and below the surface, grey opacity suffices (e.g. \citealp{2022bhatia_ssd, 2023_ssd_muram_II, 2023arXiv_ssd_muram_III}); while in the case of spectral synthesis that requires accurate temperature gradients in the stellar surface, a non-grey RT treatment is mandatory to reproduce the observations. An example of the latter includes the recent computations of the CLV by \citet{2023_muram_mps-atlas_12bins}, where they found that 12 bins were required to reproduce the solar values. 

    In our previous work, \citet{2023paperI} (hereafter \citetalias{2023paperI}), we studied 
    several strategies for the distribution of opacity bins in 1D stellar models with a solar metallicity. We found that it is possible to find optimal combinations using a small number of bins for the different spectral types studied, being a better strategy to carefully locate the bins rather than blindly increasing their number. 

    The goal of this work is twofold. First, we present our set of simulations of cool stars with the \texttt{MANCHA} code \citep{2023mancha} and compare them against the literature. 
    \texttt{MANCHA} has been extensively used to simulate the solar near-surface convection, photosphere, and low chromosphere. Some examples of these works are the study of wave propagation in sunspots \citep{2010felipe}, the study of the small-scale dynamo seeded by the Biermann battery term in the induction equation \citep{2017khomenko}, and the study of the partial ionization effects in the solar chromosphere \citep{2018khomenko}. 
    The present paper shows the first near-surface convection simulations with \texttt{MANCHA} for stars other than the Sun. 
    Second, we plan to test the conclusions from our previous work in the context of 3D MHD simulations of stellar atmospheres with solar metallicity, exploring how some of the best binning strategies found in 1D perform in the 3D simulations. 
    
    Among the references to compare with, two theses will be of special relevance to put the simulations of cool stars in context: \citet{beeck_thesis} and \citet{magic_thesis}. 
    In the associated series of papers \citep{2013beeck,2013beeck_ii}, 
    \citet{beeck_thesis} focusses in the detailed analysis of 3D hydrodynamical (HD) simulations of a set of six spectral types (ranging from F3V to M2V) with solar metallicity. Apart from HD simulations, they study the magnetic field concentrations by the stellar granulation, synthesis of three spectral lines, and CLV. \citet{magic_thesis}, and the series of papers \citep{magic2013, magic2013_ii}, shows an statistical approach, studying a grid of simulations of 217 stars with different spectral types and metallicities. 
    They describe the convection, only for purely hydrodynamical simulations of stars with effective temperature larger than 4000 K. They focus on the comparison of several spectral diagnostics computed in their 3D atmospheres versus the same parameters derived in their 1D model analogues. 

    In Sect.\ \ref{sec:method} we present the \texttt{MANCHA} code and the numerical setup for the simulations. In Sect.\ \ref{sec:description}, the results from the simulations are shown and compared with the findings of other groups. 
    In Sect.\ \ref{sec:results} the radiative energy exchange rate $Q$ is computed a posteriori in 
    one snapshot per star 
    with four opacity approaches, to describe and understand the distribution of $Q$ in the 3D cubes and to establish a reference opacity for the time evolution. 
    Finally, the impact of the different opacity approaches is characterized through the mean stratification of the temperature after time evolution.

\section{Method} \label{sec:method}
    
    \subsection{Simulations of near-surface convection with \texttt{MANCHA}} \label{subsec:convection_mancha}

   We use \texttt{MANCHA} \citep{2010felipe,2017khomenko,2018khomenko, 2023mancha}, a code that solves the time-dependent radiative MHD set of equations on a uniform 3D Cartesian grid.
   In general, the system of equations solved by the code consists of the mass continuity, the momentum conservation, the energy conservation, and the induction equation. 
   The eight primary variables evolved by the code are the mass density $\rho$, the energy per unit volume $e$, three components of the velocity $\vec{\mathrm{v}}$, and three components of the magnetic field $\vec{\mathrm{B}}$. 
   The system is closed by the equation of state (EOS) and the radiative transfer equation (RTE). The EOS 
   is used to compute the temperature $T$ and the gas pressure $p$ from the density and energy per unit volume, while the RTE 
   is used to compute the radiative energy exchange rate $Q$ that appears as a source term in the energy conservation equation. 
   The code is capable of accounting for non-ideal MHD effects (ambipolar diffusion, Ohmic heating, Hall effect) as well as for the Biermann battery \citep{2017khomenko} and thermal conduction \citep{2022navarro}.
   The design of the initial and boundary conditions in \texttt{MANCHA} is highly modular and flexible allowing for a variety of setups with different level of complexity and physical realism. 

   The spatial derivatives in \texttt{MANCHA} are computed using a 6th order finite-difference scheme, while the time-integration is performed by a 4-steps memory-saving variation of the Runge-Kutta method (see \citealp{voegler_thesis}, \citealp{2023mancha}). The code is written in modern Fortran with 3D domain decomposition using MPI library.
   
   In this preliminary study we limit the set of equations to the minimal hydrodynamic subset required for realistic simulations of stellar convection: the mass continuity, the momentum conservation, and the energy conservation equations, including the EOS and RTE:

    \begin{equation} \label{eq:mhd1}
        \frac{\partial \rho}{\partial t} + \nabla \cdot \left( \rho \vec{\mathrm{v}} \right) = \left( \frac{\partial \rho}{\partial t} \right)_{\mathrm{diff}}, 
    \end{equation}
    \begin{equation} \label{eq:mhd2}
        \frac{\partial \rho \vec{\mathrm{v}}}{\partial t} +  \nabla \cdot \left[ \rho \left( \vec{\mathrm{v}} \otimes \vec{\mathrm{v}} \right) +  p \vec{\mathrm{I}}   \right] = \rho \vec{\mathrm{g}} + \left( \frac{\partial \rho \vec{\mathrm{v}}}{\partial t} \right)_{\mathrm{diff}}, 
    \end{equation}
    \begin{equation} \label{eq:mhd3}
        \frac{\partial e}{\partial t} + \nabla \cdot \left[ \vec{\mathrm{v}} \left( e + p \right) \right] = \rho \left( \vec{\mathrm{g}} \cdot \vec{\mathrm{v}} \right) + Q + \left( \frac{\partial e}{\partial t} \right)_{\mathrm{diff}}, 
    \end{equation}
    where 
    $\vec{\mathrm{g}}$ is surface gravity, $\vec{\mathrm{I}}$ is the identity tensor, the scalar product is represented by the symbol `$\cdot$', the tensor product by `$\otimes$',  and the $\left(\dots\right)_{\mathrm{diff}}$ stands for numerical diffusivity terms. The latter are constructed following \citet[][see \citealp{2010felipe, 2023mancha}]{voegler_thesis} to mimic the form of physical diffusivity in the momentum and energy equations, while the numerical diffusivity in the mass continuity does not have a physical counterpart. The diffusivity coefficients are purely numerical and composed of three components: a constant term (applied uniformly everywhere in the domain), the hyperdiffusivity (designed to capture large local gradients), and the shock diffusivity that operates when the divergence of the velocity field is above a specified threshold. In our simulations we use only the latter two. However, for further numerical stabilization of the code we employ a 6th order filtering scheme that is described in detail in section 3.4 of \citet{2023mancha}. 

    \texttt{MANCHA} allows for flexible and user-defined boundary conditions.
    In our simulations the horizontal 
    boundary condition is periodic for all the primary variables. The top boundary is closed for vertical mass flux and energy and imposes stress-free horizontal velocities (i.e. the horizontal momentum, energy, and mass flux are set to zero at the boundary). The bottom boundary is open for the flow with implemented mass and entropy controls after \citet{voegler_thesis}.
    The entropy of the upflows is dynamically corrected to secure that the emergent flux $\mathcal{F}_{\mathrm{out}}$ of the simulation fluctuates around the specific value for every stellar type that is defined by its effective temperature $T_{\mathrm{eff}}$ (Table \ref{tab:parameters_sim_1}) as
    \begin{equation}
        \mathcal{F}_{\mathrm{out}} = \sigma T_{\mathrm{eff}}^4,
    \end{equation}
    where $\sigma$ is the Stefan-Boltzmann constant. 
    The time-scale of this correction is defined by the user, typically close to the Kelvin-Helmholtz time-scale of the domain. 
    The mass controls at the bottom boundary guarantees that the total mass in the box remains nearly constant and equal to the initial value.
    It is implemented by a correction applied to the pressure in the upflows.  
    The only important difference between our implementation and the implementation described in section 3.4.2 of \citet{voegler_thesis} is that we introduce two free parameters in the method that multiply the corrections of entropy and mass (equations 3.50 and 3.54 in \citealp{voegler_thesis}), and two more that specify the time scale at which the corrections act. 
    The parameters are tuned by hand to improve the efficiency of the corrections during the initial phase of the runs, and they are kept fixed once the simulation reaches stationary phase. 
    
    The radiative transfer equation is solved using the short characteristics method \citep{shortcharactNLTE}. The specific intensity is computed along three rays per octant, 24 rays in total, that are distributed and weighted using the quadrature A2 of \citet{1963carlson}. The density, opacity, and source function are approximated linearly on every segment along the ray. 
    As we assume validity of LTE, the source function in the grid points is computed as the Planck's function, $B$.

    To reduce the number of computations needed to solve the RTE while maintaining the quality of the bolometric results, 
    we use the opacity binning method (\citealp{norlund1982}; see \citealp{voegler_opacity_binning} and \citetalias{2023paperI} for a detailed explanation of the method) to produce the binned opacity tables. These tables are conveniently computed using opacity distribution functions \citep{1951labs_ODF, ODF1966}. The monochromatic opacity tables needed to produce  
    the ODF are computed with the \texttt{SYNSPEC} code \citep{SYNSPEC}, using its Python wrapper, \texttt{synple} \citep{synple}. 
    
    Initially the intensity is set to zero on the top boundary and to $B$ on the horizontal  
    periodic boundaries, at the bottom of the domain, and in the inner boundaries of subdomains. 
    The integration is done along the rays within each subdomain for every opacity bin. 
    To correct for inaccuracy of this initial condition in optically thin regions and for the fact that inclined rays are not fully contained in the box, the solution is iteratively 
    computed in every subdomain. 
    The results are communicated to the next subdomain in the direction of the ray propagation and compared to the local intensities there. 
    This procedure is continued until the relative error in the overlapping layer reaches $10^{-6}$. 
    
    The EOS is precomputed as a function of the temperature and density assuming instantaneous chemical equilibrium between 92 atomic and 349 molecular species. The solar composition is set after \citet{AG89} to secure consistency and to enable direct comparison with our earlier simulations of the solar convection. 
    Although it is possible to adapt the abundances to a more modern set by \citet{2021Asplund}, it would require recomputing both the opacity and the EOS tables. As the focus of our study here is not on simulations of any particular star, we leave this straightforward, but tedious step for future work. 
    
    \begin{table*}[t]
    \caption{Parameters used in the numerical setup. 
    }
    \begin{tabular}{ccccccccccc} 
    \hline
    \hline
      & $N_{\mathrm{x,y}}^2 \times N_{\mathrm{z}}$ & \begin{tabular}[c]{@{}c@{}}$dx \times dy \times dz$\\ $\mathrm{[km]}$\end{tabular} & \begin{tabular}[c]{@{}c@{}}$dt$\\ $\mathrm{[s]}$\end{tabular} &  \begin{tabular}[c]{@{}c@{}}$dt_{\mathrm{s}}$\\ $\mathrm{[s]}$\end{tabular} & \begin{tabular}[c]{@{}c@{}}$\log g$\\ $\mathrm{[cm \, s^{-2}]}$\end{tabular} & \begin{tabular}[c]{@{}c@{}}$T_{\mathrm{eff}}$\\ $\mathrm{[K]}$\end{tabular} & \begin{tabular}[c]{@{}c@{}}$\langle z \rangle_{\mathrm{\tau=1}}$\\ $\mathrm{[km]}$\end{tabular} & \begin{tabular}[c]{@{}c@{}}$\Delta x \times \Delta y \times \Delta z$\\ $\mathrm{[Mm]}$\end{tabular} & \begin{tabular}[c]{@{}c@{}}$\Delta t$\\ $\mathrm{[min]}$\end{tabular} \\ \hline
     G2V & $384^2 \times 106$ & $16 \times 16 \times 16$    & 0.05            & 30 & $4.4$ & $5780$ & 986  &  $6.1\times6.1\times1.69$ & 60 \\ 
     K0V & $384^2 \times 130$ & $10 \times 10 \times 7$     & 0.1            & 15 & $4.6$ & $4855$ & 590  & $3.8\times3.8\times0.91$   & 45 \\ 
     M2V & $432^2 \times 102$ & $3.5 \times 3.5 \times 3.6$ & $\approx 0.085$& 10 & $4.8$ & $3690$ & 256  &   $1.5\times1.5\times0.37$ & 20 \\ \hline
     \end{tabular}
     \label{tab:parameters_sim_1}
     \tablefoot{For the stellar simulations presented in this work each column displays in order the number of cells, size of cells, simulation time step, time step for saving snapshots, gravity, effective temperature, mean geometrical height of $\tau=1$ (with the zero point of the geometrical height at the bottom of the domain), total size of the domain, and total time of stationary convection}
     \end{table*}

    \subsection{Numerical setup} \label{subsec:setup}

    We use the \texttt{MANCHA} code to compute realistic time-dependent 3D radiative-hydrodynamic simulations of a sample of main sequence cool stars. In this preliminary study we limit ourselves to stars of solar metallicity and to three spectral types (G2V, K0V, and M2V) defined by the values of effective temperature and surface gravity listed in Table\ \ref{tab:parameters_sim_1} (fifth and sixth columns). The F3V star studied in \citetalias{2023paperI} was not included in the present set because the simulation was not numerically stable enough due to the formation and propagation of strong shocks. We will explore this problem in the future using higher domains and other boundary conditions at the top.

    The size of the simulation box and its spatial discretization are customized for each of the three spectral types following \citet[][second and eighth columns in Tab.\ \ref{tab:parameters_sim_1}]{2013beeck}. The vertical size of the box is chosen to incorporate around five pressure scale heights (at least two above the geometrical heights where $\tau=1$), and the horizontal size to ensure that
    there are between 15 and 20 granules within the domain. 
    The selected size of computational box minimizes the effect of the horizontal periodic boundary conditions, and of the top and bottom boundaries in the region that is our primary interest.
    The computational grid is uniform in all directions. 
    The vertical size of a grid cell is selected to be smaller than a fifth of the smallest pressure scale height (see figure 3.8 in \citealp{beeck_thesis}). In our case that is $16 \, \mathrm{km}$ for the G2V star, $7 \, \mathrm{km}$ for the K0V and $3.6 \, \mathrm{km}$ for the M2V.  The horizontal size of every grid cell is $16 \, \mathrm{km}$, $10 \, \mathrm{km}$, and $3.5 \, \mathrm{km}$ for the G2V, K0V, and M2V star, respectively. 

    The time step in the simulations (third column in Tab.\ \ref{tab:parameters_sim_1}) is set dynamically by the code 
    as the minimum of the diffusive and advective time steps and a user-defined value 
    to secure numerical stability (see section 3.1.1 in \citealp{2023mancha}). The frequency of saving snapshots (fourth column in Tab.\ \ref{tab:parameters_sim_1}) is selected to fairly sample the convection taking into account the granulation lifetime (see table 4.2 in \citealp{beeck_thesis}) and avoiding to be in phase with the stellar pulsation. 
    
    To initiate the simulations we first construct equilibrium 1D models as described in \citetalias{2023paperI}. 
    The 1D models are replicated along the horizontal direction to fill the 3D cube and a small random perturbation is added to the density and energy to seed the convection. 

    Starting from these initial snapshots, the code solves the HD equations and computes the time evolution of the convective plasma. In the initial phase we use an approximate Rosseland-mean grey opacity. At first the atmosphere experiences an instability due to the sudden radiative losses computed in the code. The random seeds start to grow and form convective flows that reach the bottom boundary of the domain after around 10-20 minutes of stellar time depending on the spectral type. Only then the energy control mechanism at the bottom boundary becomes effective. The simulation is then evolved until the outgoing stellar flux stabilizes around the prescribed value that corresponds to the effective temperature that is targeted (see App. \ref{app:thermal_relaxation}). Once the simulation with grey opacity is relaxed, we continue the run with non-grey opacity setups, again until relaxation.
    
    Four opacity setups are considered in this work (see App.\ \ref{app:bins}): Rosseland mean (hereafter \grey), four opacity bins (\obc) optimally distributed after  section 5.1 of \citetalias{2023paperI}, 18 bins (\obd) distributed in a checkerboard patter after section 5.3 of the same reference, and the same ODF computed with \texttt{SYNSPEC} monochromatic opacity as in \citetalias{2023paperI}. The first three setups are used to run the 3D stellar simulations. The ODF is only used to compute $Q$ in one snapshot of each of the three stars and compare it against the rates computed with the other three setups.\footnote{When three opacity setups are mentioned in the text, it refers to the grey, four-bins, and 18-bins opacities. If four are mentioned, it refers to these three setups plus ODF.}

    In all the results and figures presented in this work, the mean stratification of any variable with respect to the geometrical height $z$ is computed as the average over horizontal layers in the original grid of the cubes. In the case of the mean stratification of a quantity versus any other scale (i.e. optical depth), once the scale is computed in every column of the domain, the quantity is interpolated at the common grid of that scale, and then horizontally averaged. Finally, the results are averaged over one hour of stellar time for the G2V, 45 minutes for the K0V, and 20 minutes for the M2V star. These time ranges were selected to ensure that the same number of granules are considered in time and space for the three stars, taking into account the granulation lifetimes in table 4.2 in \citet{beeck_thesis}. 

\section{Description of the snapshots} \label{sec:description}

    The simulations of near-surface convection of the three spectral types (G2V, K0V, and M2V) are presented through figures and comparisons with the literature. The results in this section were computed using the \grey\ opacity setup (unless stated otherwise in the text), sufficient to illustrate the main differences between the spectral types. 
    The time averages were computed over the total stellar time indicated in Table \ref{tab:parameters_sim_1}; in some cases (e.g. the colour maps of quantities in slices of the domain) we show the results for the last snapshot in the time series of the three stars.  Any plot in this work referring to one snapshot uses that last snapshot of the corresponding spectral type.

    Three depth variables are used for the different comparisons and results. 
    The geometrical height $z$ is mostly used to show two-dimensional (2D) maps of the quantities in the stellar domain, because $z$
    defines the Cartesian grid used in the computations of the \texttt{MANCHA} code. The geometrical height is not useful to compare the change of a certain quantity with height between the different stars. To do these comparisons, the Rosseland optical depth $\tau$ and the number of pressure scale heights $N_p$ are used. The former is useful to put in context quantities related to the radiative transfer; the latter puts the emphasis on the dynamics of the domain. 

    The number of pressure scale heights is computed as
    \begin{equation} \label{eq:np}
        N_p = \ln \left( p/p_{\tau=1} \right),
    \end{equation}
    where $p_{\tau=1}$ is the pressure at the $\tau=1$ surface, the ratio of pressures $p/p_{\tau=1}$ is calculated in every column of the domain, and $\ln \equiv \log_\mathrm{e}$ ($\log \equiv \log_{10}$)\footnote{For the comparisons against \citet{beeck_thesis} take into account that they use $\log \left( p/p_{\tau=1} \right)$ instead of $\ln \left( p/p_{\tau=1} \right)$.}. 
    This definition of $N_p$ can be understood through the hydrostatical equilibrium for an ideal gas with constant temperature 
    \begin{equation}
        \ln \left( p/p_{\tau=1} \right) = -\frac{z}{H_p},
    \end{equation}
    where the pressure scale height is $H_p := P / (\rho g)$ and the reference in the geometrical height is put at $\tau=1$. With this reference, $N_p$ is zero at $\tau=1$ and has positive (negative) values in the stellar interior (atmosphere).
    
    \subsection{General properties} \label{subsec:general_prop}

    \begin{figure*}
        \centering
        \includegraphics[width=17.6cm]{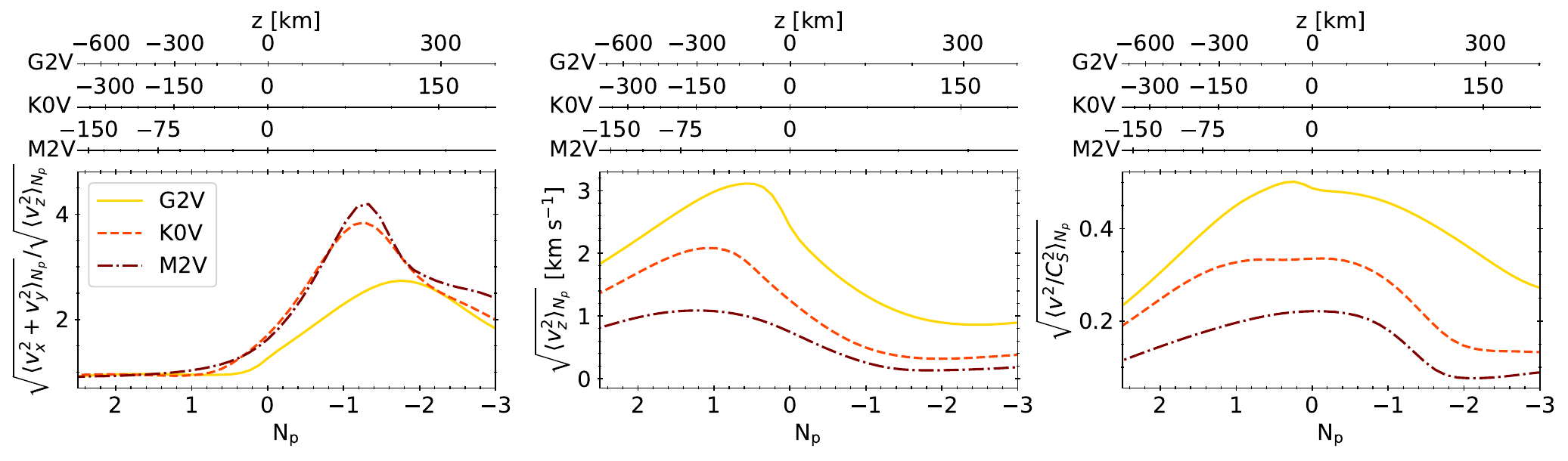}
        \caption{Mean properties of the velocity fields of the three stellar simulations (G2V is the solid yellow line, K0V is the dashed orange line, and M2V is the dot-dashed brown line) 
        averaged over time and surfaces of constant number of pressure scale heights $N_\mathrm{p}$ (Eq. \ref{eq:np}) versus $N_\mathrm{p}$. 
        Left panel: ratio of RMS horizontal and RMS vertical velocities. Middle panel: RMS vertical velocity. Right panel: Mach number. Three top axes on each column: geometrical height for the three stars. 
        }
        \label{fig:vel_stat1}
    \end{figure*}
    
    \begin{figure*}
        \centering
        \includegraphics[width=17.6cm]{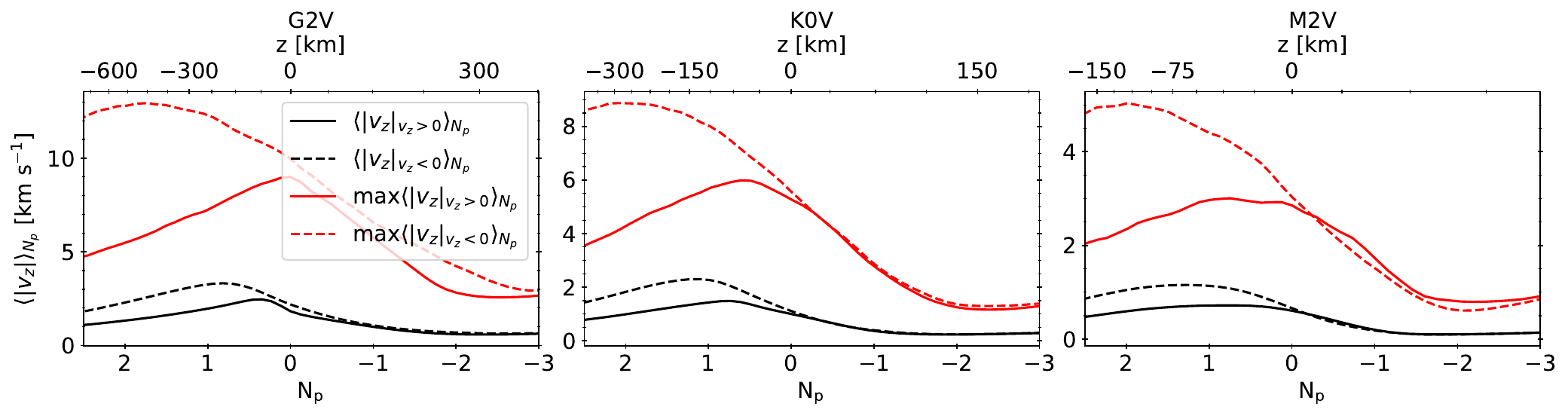}
        \caption{Mean (black) and maximum (red) vertical velocity of upflows (solid line) and downflows (dashed) averaged over surfaces of the constant number of pressure scale heights $N_\mathrm{p}$ and time for the three stellar simulations (G2V is in the left panel, K0V is in the middle, and M2V is in the right) versus $N_\mathrm{p}$.
        The top axis of each panel shows the geometrical height. 
        }
        \label{fig:vel_stat2}
    \end{figure*}

    \begin{figure}
        \centering
        \includegraphics[width=8.8cm]{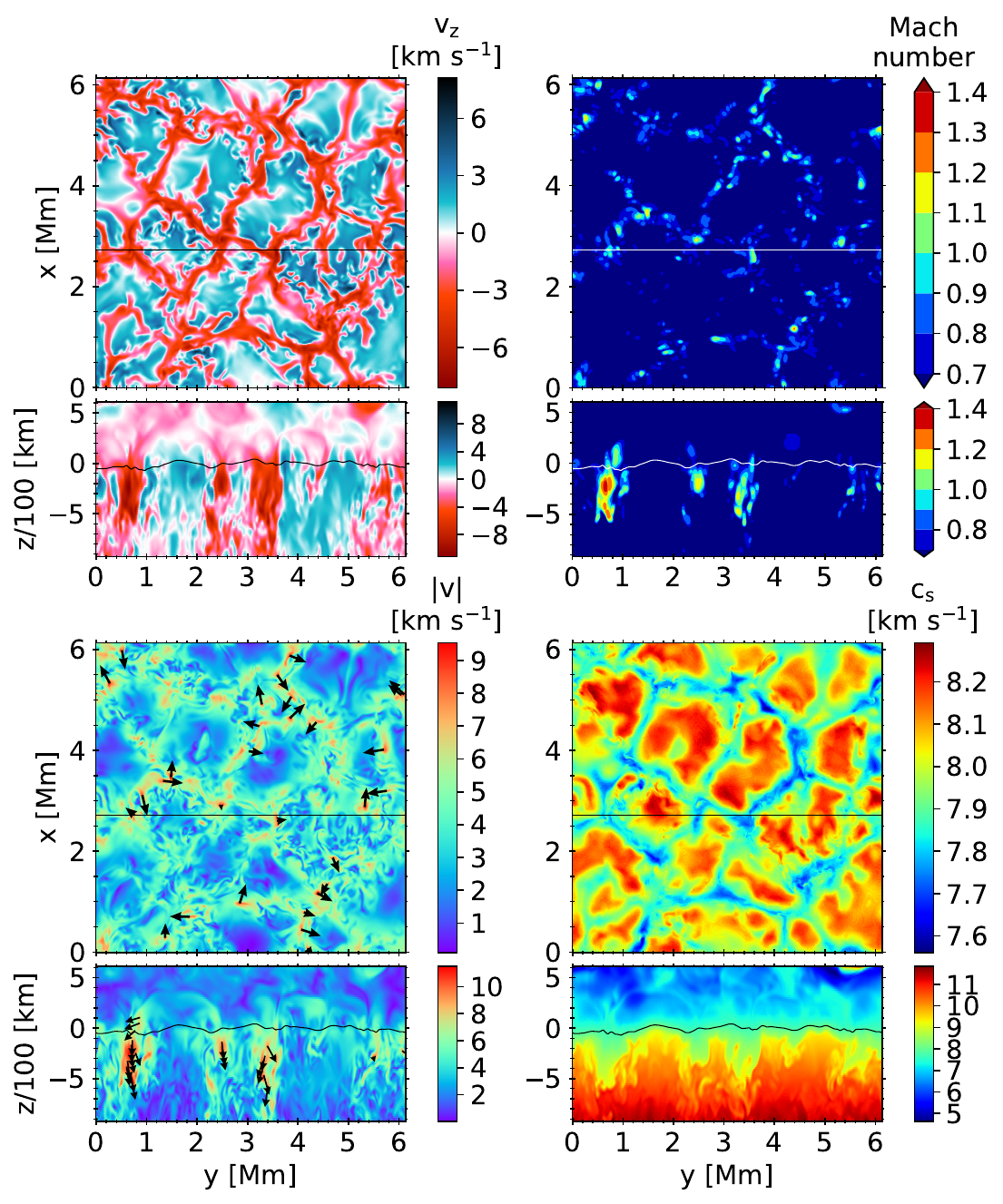}
        \caption{Colour maps of the vertical velocity (top left quadrant), the Mach number (top right quadrant), the modulus of velocity (bottom left quadrant), and the sound speed (bottom right quadrant) of the last snapshot of the time series of the G2V star.
        Each quadrant shows the $\tau=1$ surface (top panel) and a vertical cut at $x=2.72 \, \mathrm{Mm}$ (bottom panel). The location of the $\tau=1$ surface is shown with a solid curve in the bottom panel of every quadrant and the location of the vertical cut is shown as a solid horizontal line in the top panel. The values of the mapped quantities are shown in the corresponding colour-bar positioned next to each panel. 
        In the top right quadrant 
        the green, yellow, orange, and red colours correspond to supersonic Mach numbers. 
        The velocity direction at the supersonic flows is indicated by black arrows in the panels of the bottom left quadrant (projected in the $\tau=1$ surface and the vertical plane).
        }
        \label{fig:mach_g2v}
    \end{figure}
    
    Figure \ref{fig:vel_stat1} shows some mean properties of the velocity field of the three spectral types against the number of pressure scale heights. 
    The left panel of the figure shows the mean stratification of the ratio of the root mean square (RMS) horizontal and RMS vertical velocity. This quantity offers an idea of the 3D structure of the velocity. Below the surface, the vertical component of the velocity dominates the motion of the plasma, the ratio is lower than $\sqrt{2}$. Consistent with \citet{2013beeck}, we find the ratio to be close to unity for $N_\mathrm{p}>1$. For $N_\mathrm{p}<1$, the horizontal velocity becomes larger than the vertical velocity and the ratio hits its maximum at $N_\mathrm{p} \simeq -1$ for the K0V and M2V stars, and $N_p \simeq -2$ for the G2V. 
    The amplitude of the peak is around $2.8$ for the G2V star, $3.8$ for the K0V star, and $4.4$ for the M2V star.
    These amplitudes and locations of the peaks are close to the ones found in \citet{2013beeck} for G2V and K0V stars, while our simulation of the M2V star seems to have its peak located one pressure scale higher in the atmosphere with a larger amplitude.
    
    The mean stratification of the RMS vertical velocity is shown in the middle panel of Fig.\ \ref{fig:vel_stat1}. It peaks close to the stellar surface, with larger amplitude for the G2V star, indicating a more vigorous convection. The amplitude is around $3$ km s$^{-1}$ for the G2V star, $2$ km s$^{-1}$ for the K0V star, and $1$ km s$^{-1}$ for the M2V star. These values are consistent with the values shown in figure 3.19 from \citet{magic_thesis} and figure 6 from \citet{2013beeck}, with the exception of the M2V star, for which \citet{2013beeck} finds a lower amplitude. More details about the vertical velocity can be found in Fig.\ \ref{fig:vel_stat2}, where the mean and the maximum values of the vertical velocity of upflows and downflows is shown. In all stars, the velocity of upflows and downflows peak close to the surface of the star, having the downflows larger velocities than the upflows in the subsurface layers, and both downflows and upflows a similar velocity higher up. This description is close to the one in \citet[][see the right panel in their figure 7]{2013beeck}. 

    \begin{figure*}
        \centering
        \includegraphics[width=17.6cm]{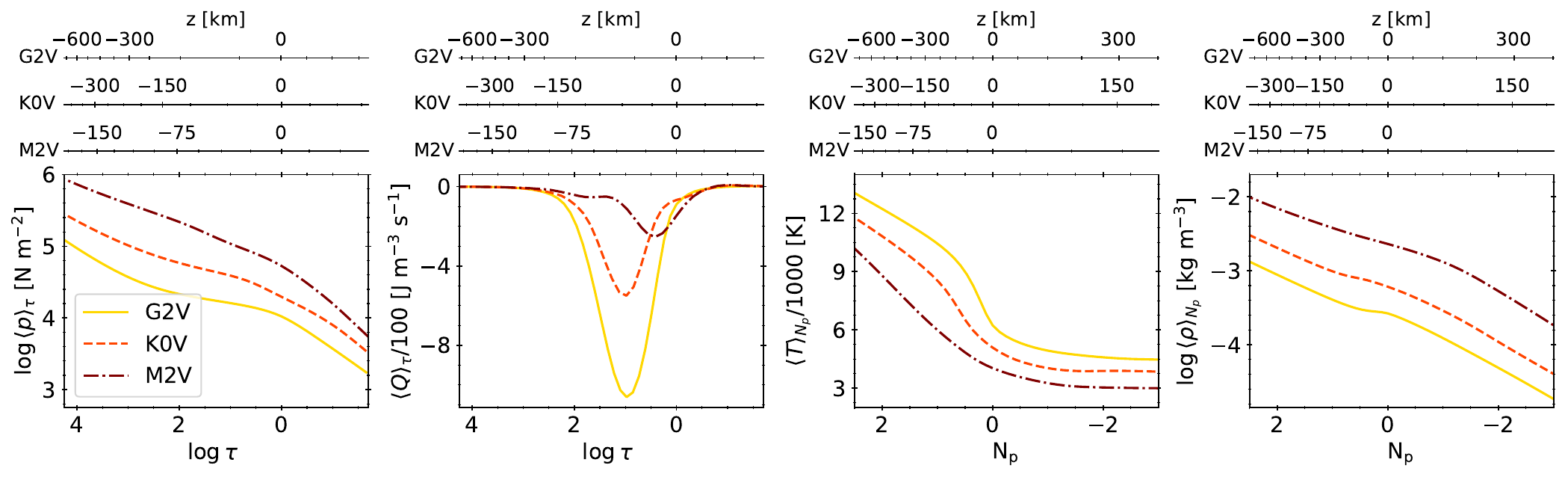}
        \caption{Logarithm of the pressure (left panel), radiative energy exchange rate (second), temperature (third), and logarithm of the density (right)  averaged over time and Rosseland iso-$\tau$ surfaces (two left panels) or surfaces of a constant number of pressure scale heights (two right panels). The colours and line styles have the same meaning as in Fig. \ref{fig:vel_stat1}. Three top axes of each column: geometrical height for the three stars.}
        \label{fig:1d_prop}
    \end{figure*}

    The mean stratification of the Mach number is shown in the right panel of Fig.\ \ref{fig:vel_stat1}, depicting a progressively smaller Mach number from G2V to M2V star that peaks close to the stellar surface and is subsonic for the three stars (with the values of the amplitude close to those shown in the right panel of figure 6 from \citealp{2013beeck}). 
    Although in the mean stratification there is no glimpse of supersonic flows, they appear in the 3D velocity field of our G2V star simulation. As shown in the literature, supersonic flows are 
    an observed \citep{2009BellotRubio} and simulated \citep{1998SteinNordlund} phenomenon in the solar photosphere,  with a remarkably good match between simulations and observations \citep{2011Vitas_supersonic}. 
    Figure \ref{fig:mach_g2v} shows some structures with Mach number higher than one in the 3D velocity field of the last snapshot in the time series of the G2V star. 
    In the figure, the vertical velocity, Mach number, modulus of the velocity, velocity field, and sound speed show that the supersonic flows appear in the strongest downflows mainly in the intergranular lanes, sometimes in the edge of the granules above the surface. This is consistent with the description from \citet{1998SteinNordlund}, as it is the fact that the supersonic flows close to the surface ($\log \tau \in [-1,1]$) cover around 3\%-3.5\% of the total area of the domain. Around $\lesssim 0.01\%$ of the cells in each snapshot in the G2V time series show supersonic flows, with Mach number up to $\simeq 1.8$. 
    Some supersonic flows appear too in the K0V star simulation, but only in the strongest downflows in the intergranular lanes (approximately one in every $10^6$ cells has supersonic flow, with Mach number up to $\simeq 1.1$). There is no trace of supersonic flows in the M2V simulation. 

    In Figure \ref{fig:1d_prop} we show other properties of the mean stratification, such as pressure $p$, radiative energy exchange rate $Q$, temperature $T$, and density $\rho$. 
    In the left panel of the figure, the pressure averaged over surfaces of the constant Rosseland $\tau$ is shown. As described in \citet{2013beeck}, in the surface the logarithm of the pressure essentially depends linearly on the logarithm of the optical depth, while in the deeper layers the convection determines the pressure profiles. 
    
    The second panel of Fig.\ \ref{fig:1d_prop} shows $Q$ averaged over surfaces of the constant Rosseland $\tau$. 
    The profiles are similar to those shown for 1D models in \citetalias{2023paperI}. The $Q$ rate of the G2V and K0V stars shows one cooling lobe ($Q<0$) close to $\log \tau = 1$ and one heating lobe ($Q>0$) over the surface, with negligible amplitude with respect to the cooling (the heating lobe is barely detectable in this figure, see also Fig.~\ref{fig:q_kappa_n_G_M}).
    The $Q$ rate of the M2V star has a different shape compared to the other spectral types. The decrease of $Q$ in the subsurface layers is interrupted around $\log \tau=1.7$, where the $Q$ rate is almost constant for about 50 km (hereafter feature A; see Section \ref{subsec:Q_M2V} for more details). As it is subsequently shown in Section \ref{subsec:Q}, the relative amplitude of the heating with respect to the cooling increases for lower effective temperature, owing to blanketing coming from molecular lines (the effect is negligible with grey opacity). This effect, evident in the M2V star but very small in the two hotter stars, is consistent with the systematic study using 1D models from \citet{2008marcs}, where they show that the backwarming (consequence of the line blanketing) increases significantly for $T_{\mathrm{eff}}<4000$ K, thanks to the line blocking of TiO and H$_2$O, and other molecules that involve $\alpha$-elements \citep{2012ambre}. 
    In our studied set of stars, the amplitude of the cooling decreases for lower effective temperature, consistent with the results from section 3.2.8 of \citet{magic_thesis}. The minimum of $Q$ in the M2V star is located higher in the atmosphere than in the other stars, around $\log \tau = 0.4$.  
    This result is in odds with the trend shown in figure 3.30 from \citet{magic_thesis}, although in this study they do not include stars with $T_{\mathrm{eff}}<4000$ K.
    
    In the third panel of the Figure \ref{fig:1d_prop}, the temperature versus the number of pressure scale heights of the G2V and K0V stars shows a steep gradient at the surface of the star, while the M2V presents a smoother gradient. These temperature gradients are consistent with the ones found in the left panel of figure 10 from \citet{2013beeck}. 
    The right panel in Fig. \ref{fig:1d_prop} shows the logarithm of the density versus the number of pressure scale heights. The logarithm of the density depends almost linearly on the number of pressure scale heights, except for the layers close to the surface. 
    
    To complete the picture outlined with the differences in the vertical stratifications between the spectral types, the appearance of the actual granulation is presented. Figure \ref{fig:stars_2d} shows the temperature, vertical velocity, density, and radiative energy exchange rate $Q$ in the $\tau=1$ surface and a vertical cut in the middle of the last snapshot of the time series for the three stars. 
    Consistent with \citet[][see his figure 4.5 and the beginning of his section 3.1.1]{beeck_thesis} and \citet[][see his figure 3.11]{magic_thesis}, we find that the granulation of the three spectral types is qualitatively different. 
    The size of the granules gets progressively smaller from G2V to M2V spectral type (around $1$ Mm of diameter for the G2V star, 0.5 Mm for the K0V, and 0.25 Mm for the M2V) and their shape goes from smooth granules with rounded edges in the G2V star to more irregular granules with smaller sub-structures in the M2V. This variation in the appearance of the granulation is also shown in the study of the fractal dimension of the granules in section 4.3 from \citet{magic_thesis}, where they found that for lower effective temperatures and larger surface gravities, the granules show more irregular perimeters. We also see how the $\tau=1$ surface is more corrugated in the G2V star, with a standard deviation of $z$ of around 30 km, while the standard deviation is 6 km for the K0V and 1 km for the M2V\footnote{These measures of the corrugation of the $\tau=1$ surfaces are only to illustrate the difference between the spectral types. Since these values are lower than the cell height in the snapshots for the K0V and M2V, to achieve more accuracy one should increase the spatial resolution in the simulations.}. The change of the corrugation of the iso-$\tau$ surfaces for different spectral types is also shown in figure 3 from \citet{2013beeck_ii} and section 4.4.1 from \citet{magic_thesis}.
    
    The top row in the Figure shows the temperature in the $\tau=1$ surface, revealing a surface temperature contrast that decreases for decreasing effective temperature. The exact values of the stratification of the temperature contrast are given in Fig.\ \ref{fig:tem_contrast}, averaged over surfaces of the constant Rosseland optical depth and time for the three spectral types. From M2V to G2V the profile of the temperature contrast is sharper, with larger amplitude and the maximum closer to the surface. This is consistent with the top panel of figure 12 in \citet{2013beeck}.
    
    The second row in Fig.\ \ref{fig:stars_2d} shows the temperature in a vertical cut, from which more corrugated isotherms are seen for larger effective temperatures. The mean Rosseland optical depth of the isotherm that corresponds to the effective temperature $\langle \tau \rangle_{T=T\mathrm{_{eff}}}$ is around 0.6, 0.7, and 0.5 in the time series of the G2V, K0V, and M2V stars, respectively. These values are close to the value $\tau=2/3$ predicted in the Eddington-Barbier approximation for the stellar flux. 
    
    The third and fourth rows in Fig.\ \ref{fig:stars_2d} show a progressively smaller velocity from G2V to M2V spectral type, indicating a less vigorous convection, as already shown in the middle panel of Fig.\ \ref{fig:vel_stat1}. 
    The fifth and sixth rows in Fig.\ \ref{fig:stars_2d} show the density, that gradually increases in the stellar surface from the G2V to the M2V. 
    The vertical cuts of the vertical velocity and density also reveal a more complex 3D structure with more corrugated iso-surfaces for higher effective temperatures. This increase of the complexity of the 3D structures for higher effective temperature is consistent with the analysis presented in section 3.2 and figure 3 in \citealp{2013beeck_ii}. 
    
    Finally, the two bottom rows in Fig.\ \ref{fig:stars_2d} show $Q$ in the $\tau=1$ surface and the vertical cut. The cooling ($Q<0$) mainly concentrates right below the $\tau=1$ surface. We find large values of the heating ($Q>0$) in the downflows in the intergranular lanes, although these are not present in the mean stratification (second panel of Fig.\ \ref{fig:1d_prop}) since the cooling in these heights contributes more to the average. 
    
    To better understand how the surface is cooled by radiation, the same vertical cut showing $Q$ versus two height coordinates ($\log \tau$ and $N_p$) is shown in Figure \ref{fig:compareQ_tau_npre}. 
    In the top row, where $Q$ is shown versus the optical depth, we can see how the bulk of the cooling happens in $\log \tau \in [0, 2]$ in the three stars. The picture is different if we look at $Q$ in terms of $N_p$ (middle row), where it is clear that this cooling layer gets relatively thicker for lower effective temperature. The bottom row of Fig. \ref{fig:compareQ_tau_npre} shows the temperature versus $N_p$, 
    evincing a sharper drop of temperature with height for larger effective temperatures. 
    The bottom row also displays lines with constant $\log \tau$, showing that for a fixed range of optical depths, the width of the range of $N_p$ increases for decreasing effective temperature. 
    
    The behaviour of $Q$ and $T$ versus $N_p$ depicts an atmosphere with a relatively sharp surface for the G2V star (the photons from the surface come from a shallow optically thin layer), and a more smeared surface in the case of the M2V star (the photons escape from different layers that are gradually optically thinner). 
    The smeared surface, together with the weaker convection and the lower temperature contrast, produce less visible granular surfaces in the M2V star, phenomenon known as veiled granulation \citep{nordlund_dravins_1990, 2013beeck, 2013beeck_ii}. 

    \subsection{Radiative energy exchange rate of the M2V star} \label{subsec:Q_M2V}
    
\begin{figure*}[h]
        \centering
        \includegraphics[width=17.6cm]{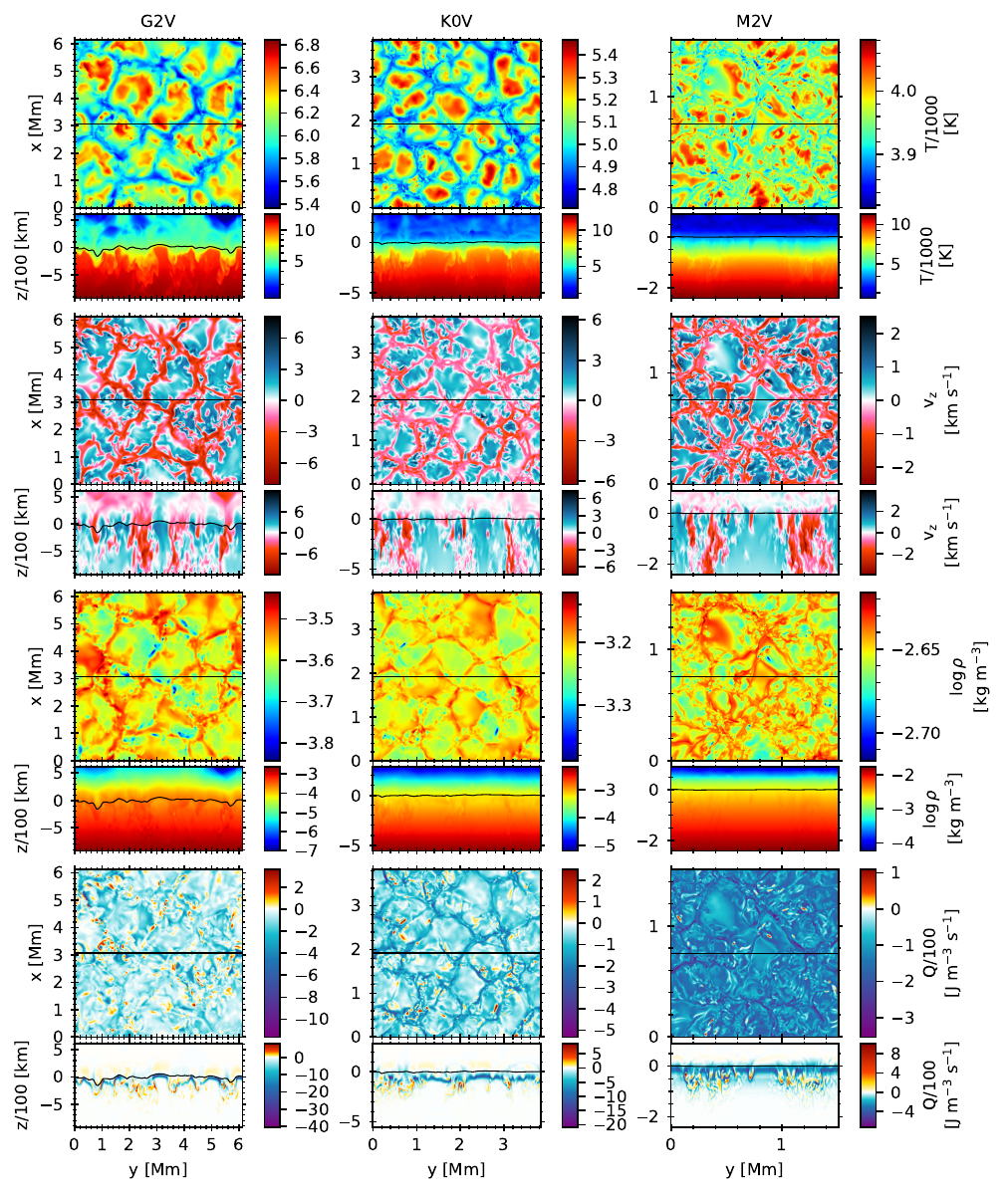}
        \caption{Temperature (top pair of rows), vertical velocity (second pair), logarithm of density (third), and $Q$ rate (bottom) of the last snapshot of the time series of the G2V (left column), K0V (middle), and M2V (right) stars. The values of the quantities shown in the colour maps are indicated by the corresponding colour-bars. The upper panel of each pair of rows shows the surface at $\tau=1$; the lower panel, a vertical cut at the middle of the snapshot. Similarly to Fig.\ \ref{fig:mach_g2v}, the location of the iso-$\tau$ surface and the vertical cut are shown as solid curves. }
        \label{fig:stars_2d}
    \end{figure*}

\begin{figure}[h!]
        \centering
        \includegraphics[width=8.8cm]{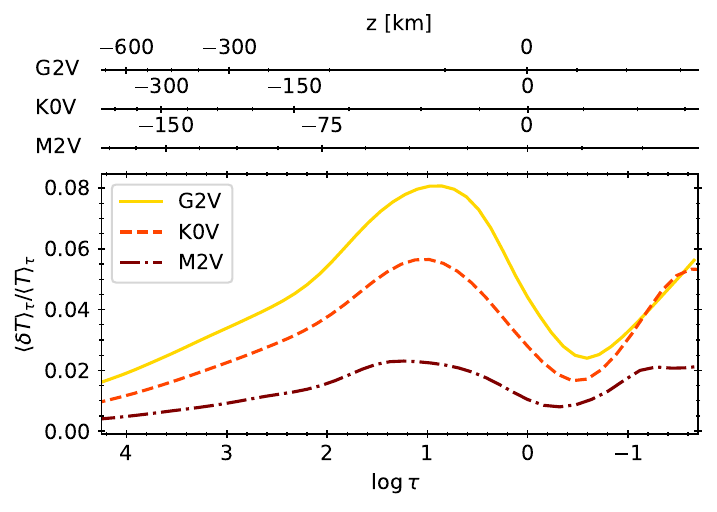}
        \caption{Temperature contrast averaged over Rosseland iso-$\tau$ surfaces and time for the three spectral types. The colours and line styles have the same meaning as in Fig. \ref{fig:vel_stat1}. The contrast is computed as the standard deviation of the temperature divided by its mean.}
        \label{fig:tem_contrast}
    \end{figure}

    The mean stratification of $Q$ of the M2V star is significantly different from the $Q$ from the G2V and the K0V stars (see second panel of Fig.\ \ref{fig:1d_prop}, bottom row of Fig.\ \ref{fig:stars_2d} and second row of Fig.\ \ref{fig:q_kappa_n_G_M}). First, while the other stars show one smooth minimum for $Q$, 
    this minimum is interrupted with 
    $Q \approx \mathrm{constant}$ at $N_\mathrm{p} \in [0.9,1.3]$ in the M2V atmosphere (feature A in the right panel of the second row Fig.\ \ref{fig:q_kappa_n_G_M}). 
    Second, the heating (feature B in in the same panel) over the surface of the M2V is significantly larger than the heating in the other stars if non-grey opacities are used (see Sect.\ \ref{subsec:Q}). The latter is explained by the blanketing produced by molecular lines (see \citealp{2008marcs}), but the former is less obvious. An answer to why $Q \approx \mathrm{constant}$ in feature A would require a detailed analysis on the feedback between the formation of the stellar temperature gradient 
    and the opacity through 1D modelling. This is out of the scope of this work, but we present a deeper description of the problem to guide its understanding.

\begin{figure*}[bh!]
        \centering
        \includegraphics[width=17.6cm]{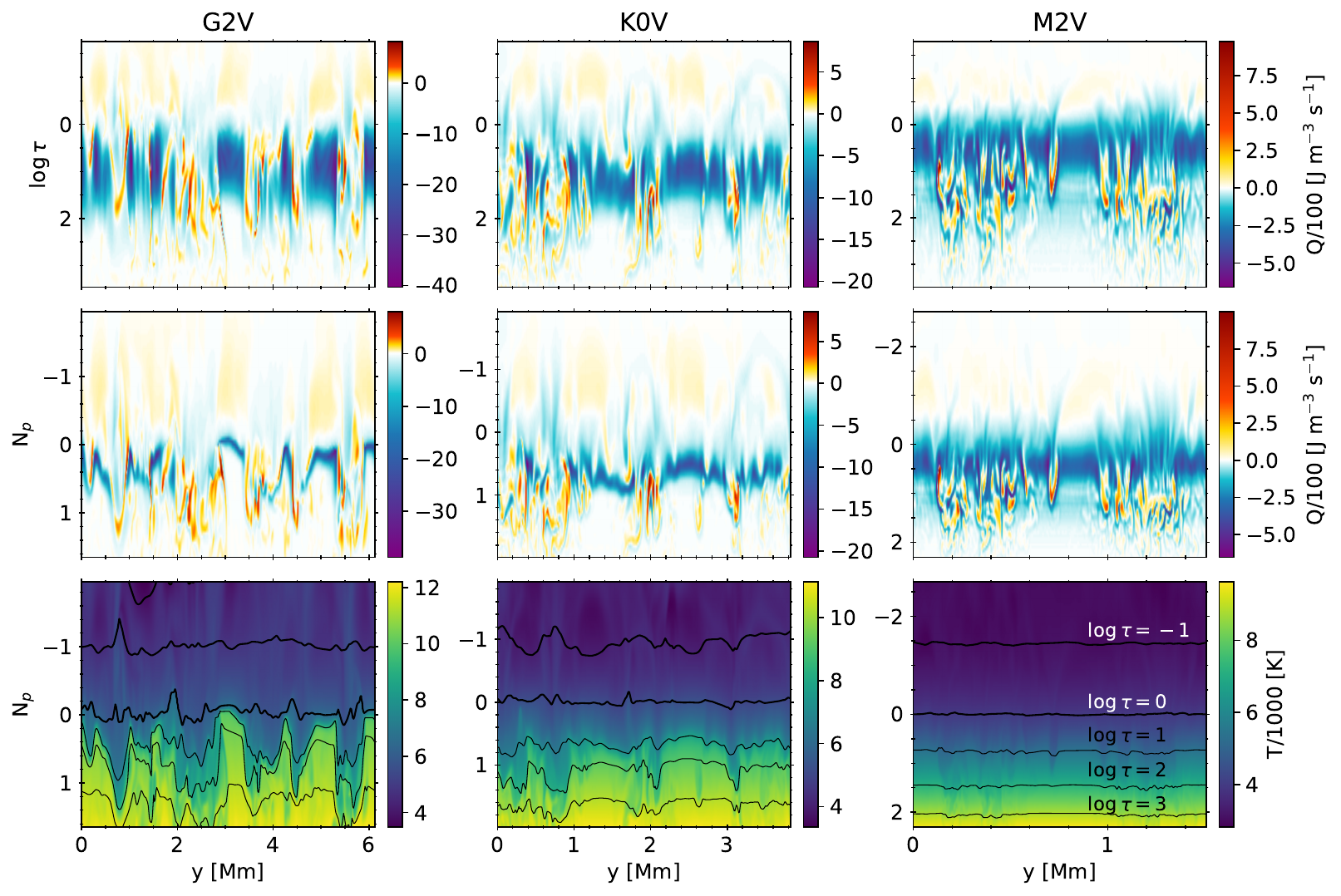}
        \caption{Radiative energy exchange rate $Q$ (top and middle row) and temperature (bottom row) for the same vertical cut as in Fig.\ \ref{fig:stars_2d} in the last snapshot of the time series of the three spectral types (different columns). The vertical cut is shown with the geometrical coordinate $y$ in the horizontal axis, and the logarithm of the Rosseland optical depth (top row) and the number of pressure scale heights (middle and bottom rows) in the vertical axis. Both depth scales show the same vertical extension. In the bottom row, the iso-$\tau$ surfaces are shown in thick solid black ($\log \tau \le 0$) and thin solid black ($\log \tau > 0$) lines, with $\Delta \log \tau = 1$. }
        \label{fig:compareQ_tau_npre}
    \end{figure*}

\begin{figure*}
        \sidecaption
        \includegraphics[width=12cm]{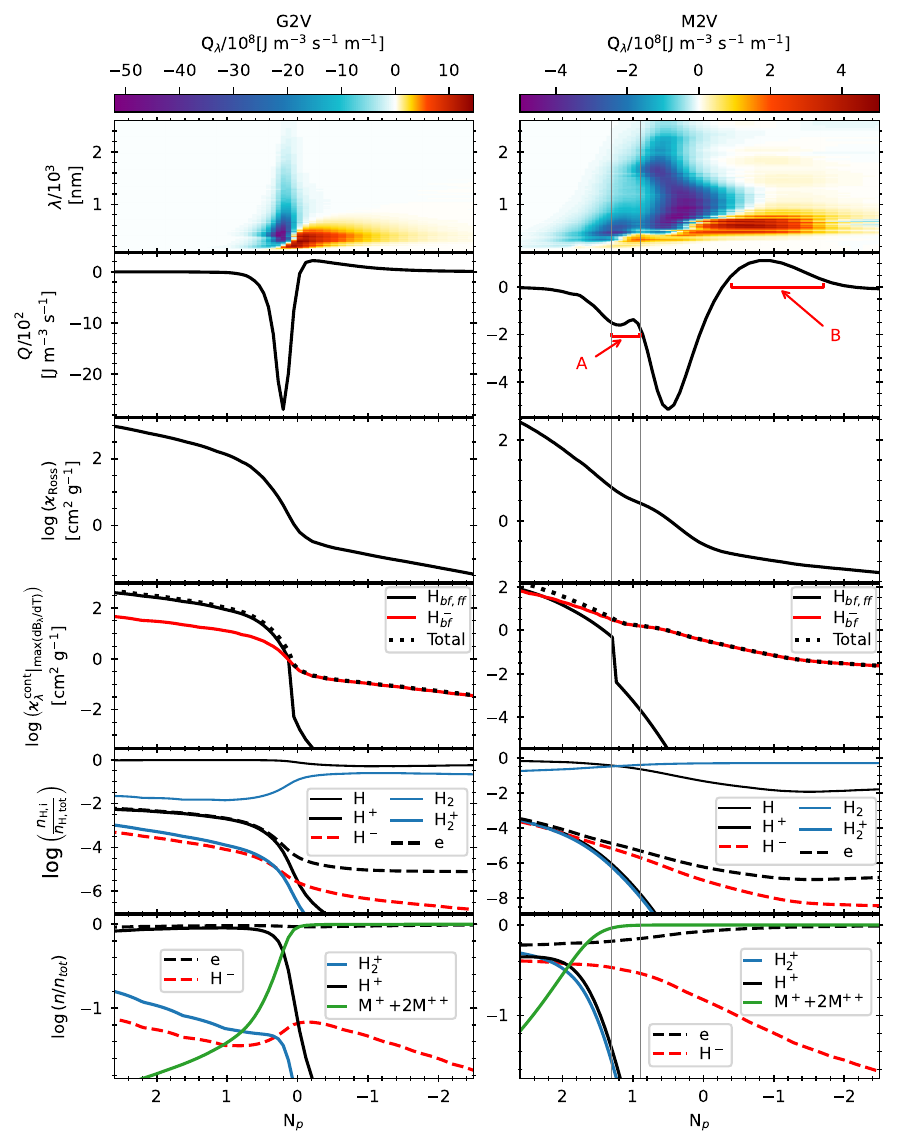}
        \caption{Variation with $N_\mathrm{p}$ of the radiative energy exchange rate, opacity, nuclei conservation, and charge conservation for the G2V (left column) and M2V (right) stars for the last snapshot in their time series. 
        The top row shows the 2D colour-map of the monochromatic $Q_\lambda$ computed with the ODF withing each $\lambda$-step ($\sum_{i,j} Q_{i,j} \omega_j$ in the notation of \citetalias{2023paperI}). The vertical axes of this row show the wavelength. The corresponding colour-bar located over each panel ranges from $\min Q_\lambda$ to $\max Q_\lambda$, and displays the negative $Q_\lambda$ (i.e. cooling) in purple-blue tones and the positive $Q_\lambda$ (i.e. heating) in red-yellow tones. 
        The second and third rows show the bolometric $Q$ and logarithm of Rosseland opacity, respectively. 
        The fourth row shows the logarithm of the monochromatic opacity at the wavelength where the derivative of the Planck function $dB_\lambda / dT$ is maximum for the total continuum opacity (black dots) and its two main contributors, bound-free plus free-free transitions for neutral hydrogen (solid black line) and bound-free transitions for H$^-$ (solid red line). The discontinuity in the opacity of neutral hydrogen corresponds to the shift through its first ionization edge. 
        The fifth row shows the nuclei conservation of hydrogen as the logarithm of 
        the ratio of the number density of each H specie and the total number of H nuclei. The same ratio is also shown for the electrons (dashed black line). 
        The bottom row shows the charge conservation as the logarithm of the 
        ratio of the number density of the charges from each specie and the total number of negative charges. The species are specified in the legend. The electrons coming from ionization of atoms with atomic number $Z>1$ (i.e. metals M) are labelled as $\mathrm{M}^+ + 2\mathrm{M}^{++}$. 
        In all the panels of the right column, the range $N_\mathrm{p} \in [0.9, 1.3]$ is marked by two vertical grey lines. }
        \label{fig:q_kappa_n_G_M}
    \end{figure*}

\begin{figure*}
        \centering
        \includegraphics[width=16.cm]{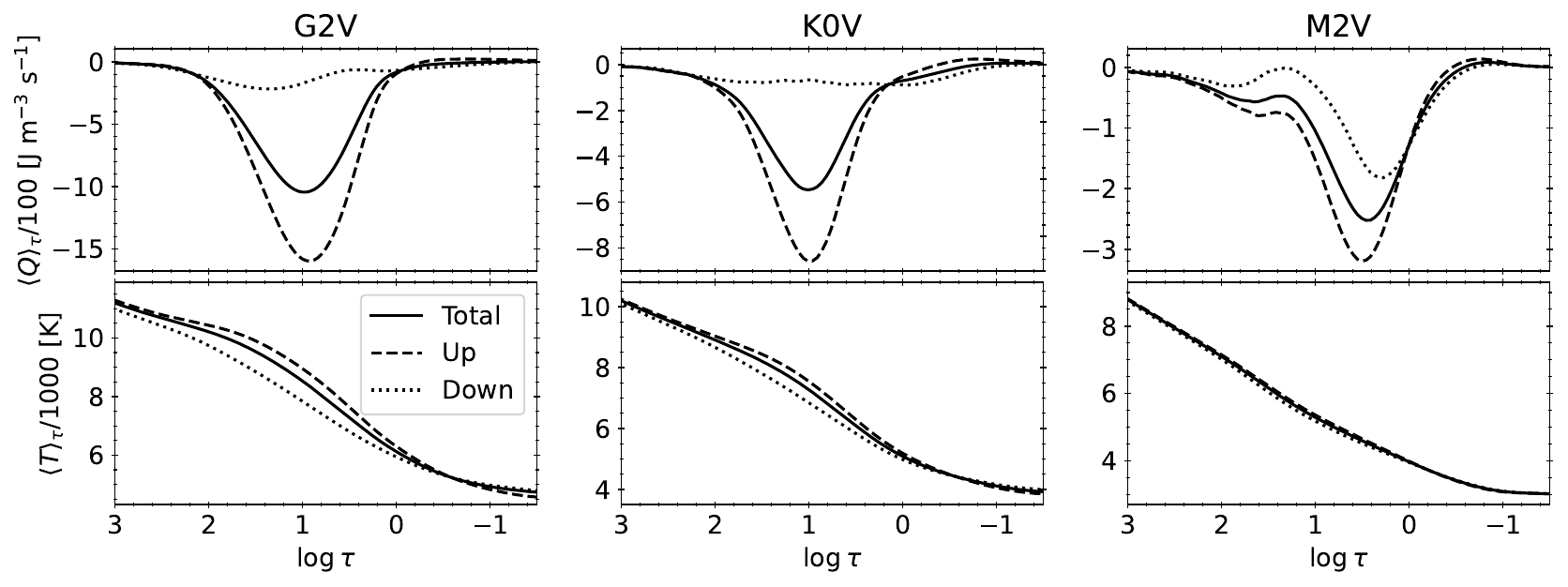}
        \caption{Mean of the radiative energy exchange rate $Q$ (top row) and temperature (bottom) over the iso-$\tau$ surfaces for the entire domain (solid line) and separated for upflows (dotted) and downflows (dashed). The results are shown for the last snapshot of the time series of the G2V (left column), K0V (middle), and M2V (right) star.
        }
        \label{fig:up_down}
    \end{figure*}

    Although the large values of heating in the downflows of the intergranular lanes may suggest that the mean stratification of $Q$ comes from an interplay of upflows and downflows (top row of Fig.\ \ref{fig:compareQ_tau_npre}), 
    the feature A appears in 1D computations too (see \citetalias{2023paperI} and second row of Fig.\ \ref{fig:q_kappa_n_G_M}). This adds to the fact that the nearly constant $Q$ in $N_\mathrm{p} \in [0.9,1.3]$ appears in downflows as well as upflows, as shown in Fig.\ \ref{fig:up_down}. 
    Thus, the $Q$ in feature A is driven by the mean stratification, rather than being a 3D effect. 

    Since feature A is reproduced by 1D models, we take the mean stratification of the last 3D snapshot of the M2V and G2V (as a reference) time series and check the $Q$ rate, the opacity and the equation of state with height (Fig. \ref{fig:q_kappa_n_G_M}). 
    The $Q$ rates are computed with the ODFs (one with $T$, $\rho$ grid adapted for the G2V and K0V star and another for the M2V star) computed with \texttt{SYNSPEC} \citep{SYNSPEC} opacities and the RTE solver from \citetalias{2023paperI}. 
    The second row of the Figure \ref{fig:q_kappa_n_G_M} shows the bolometric $Q$. In the case of the G2V, the cooling $Q<0$ is mainly found at $N_\mathrm{p} \in [0,1]$ (defining a sharp surface), while in the M2V it is blurred around $N_\mathrm{p} \in [-0.3, 2]$ (smeared surface). 

    The opacity of the two stars must be related with their differences in the $Q$ rates. This is evident in the Rosseland opacity, shown in the third row of Fig.\ \ref{fig:q_kappa_n_G_M}, since going up in the atmosphere of the G2V star, the opacity decreases sharply at the surface, while in the M2V it smoothly decreases. The opacity of the M2V star displays a change of slope 
    around $N_\mathrm{p} \in [0.9, 1.3]$, where $Q$ remains almost constant.
    
    To understand which parts of the spectrum produces the feature A in $Q$, the 2D colour-map of the monochromatic $Q_\lambda$ rate computed with the ODF within each $\lambda$-step\footnote{The final contribution to the bolometric integral of each ODF-step is proportional to the length of its $\lambda$-range (check \citetalias{2023paperI} for the details of the ODF formalism).} is shown at the top row of Fig.~\ref{fig:q_kappa_n_G_M}. 
    The wavelengths that contribute to the bolometric cooling at $N_\mathrm{p} \in [0.9, 1.3]$ are mainly in the visible and in the near infrared. For those wavelengths and depths, the opacity of cool stars is dominated by the continuum opacity. Thus, we can check the components of the continuum opacity to identify the source of this change of slope. 
    The monochromatic continuum opacity (black dots) and its main components are shown at the fourth panel of Fig.\ \ref{fig:q_kappa_n_G_M}. The opacity in this figure is shown at the wavelength where the derivative of the Plank function $dB_\lambda / dT$ is maximum. It is done this way because $dB_\lambda / dT$ is the weighting   
    \clearpage
    \noindent function in the Rosseland mean (Eq. \ref{eq:harmonic_mean}). Since the continuum dominates the Rosseland mean, the wavelength  at which $dB_\lambda / dT$ is maximum is a good choice (the opacities close to the maximum of $dB_\lambda / dT$ have thus maximum weight in the average and help us to see which opacities dominate).
    The continuum is dominated by the H$^-$ bound-free transitions (red line) at $N_\mathrm{p} \in [0.9, 1.3]$, as it is expected for cool stars (see section 8 in \citealp{greyBook}). To understand  the role  of the EOS in producing the opacity from H$^-$, the conservation of H nuclei and charge are shown at the two bottom rows of Fig.~\ref{fig:q_kappa_n_G_M}.

    The fifth row of Fig.~\ref{fig:q_kappa_n_G_M} shows the conservation of H nuclei as the ratio of the number density of each H species and the total number of H nuclei. Going up in the atmosphere, in both stars the number densities of neutral and positive ions of H decrease (thin solid black and 
    thick solid black lines), while the number density of molecular hydrogen increases (thin blue line). Again, this variation happens relatively abruptly in the sharp surface of the G2V star and more smoothly in the M2V smeared surface. Another difference between the two stars is that H$_2$ (thin blue line) is way more abundant in the M2V than in the G2V, being its dominant hydrogen specie for $N_\mathrm{p}<1$ (see e.g. \citealp{1966vardya} and the section about molecule formation in chapter 3.2 from \citealp{1970mihalas}). The hydrogen negative ion (dashed red line) shows a similar behaviour in both stars, decreasing its number density with height owing to the drop of H$^+$ (thick solid black line).
    
    The bottom row of Fig.~\ref{fig:q_kappa_n_G_M} shows the charge conservation as the ratio of the number density of the charges from each specie and the total number of negative charges. In both stars the decrease of H$^-$ (dashed red line) produced by the drop of H$^+$ (solid black line) is slowed down owing to the ionization of the metals (green line), that start to contribute significantly to the electron pool (dashed black line). 
    The metals become electron donors in the sharp surface of the G2V, while this happens in a relatively 
    narrow range of $N_\mathrm{p}$ compared to the wide and 
    smeared surface of the M2V star. In that relatively narrow 
    range around $N_\mathrm{p} \in [0.9, 1.3]$ the amount of electrons coming from ionization of metals ensures that there is enough H$^-$ so the decrease of the opacity from its bound-free transitions is slowed down, staying $Q$ almost constant.

\section{Results} \label{sec:results}
    
    In the present section, we first analyse the differences between the $Q$ rates computed a posteriori with four opacity setups using only one snapshot of each of the three spectral types (Sect.\ \ref{subsec:Q}). The aim is to describe and understand the distribution of $Q$ in the 3D cubes. Moreover, through this analysis we will establish that the \obd\ setup yields results that are so close to the results obtained with the ODF that it can substitute ODF as the reference setup 
    in Sect.\ \ref{subsec:Temperature}. In that section, simulations run with three opacity setups are used to evaluate 
    the impact of the different opacities on the temperature stratification.

    \subsection{Effect of the binning on $Q$: The one-snapshot case} \label{subsec:Q}

    The bolometric radiative energy exchange rate $Q$ is studied in detail because it is the only term in the (M)HD set of equations (Eqs. \ref{eq:mhd1}-\ref{eq:mhd3}) that accounts for interactions between the radiation and the matter. The $Q$ rate (J m$^{-3}$ s$^{-1}$) 
    is particularly sensitive to the low photosphere which is the primary target of our simulations and where our LTE assumption is valid and consistent with the computations of the opacity table. In contrast to that, the ratio $Q/\rho$ emphasizes the effects of the radiation and the opacity strategy choice in the regions higher up in the atmosphere, with much lower density (cf. figure 4.10 in \citealp{voegler_thesis}).

    \begin{figure}[h]
        \centering
        \includegraphics[width=8.8cm]{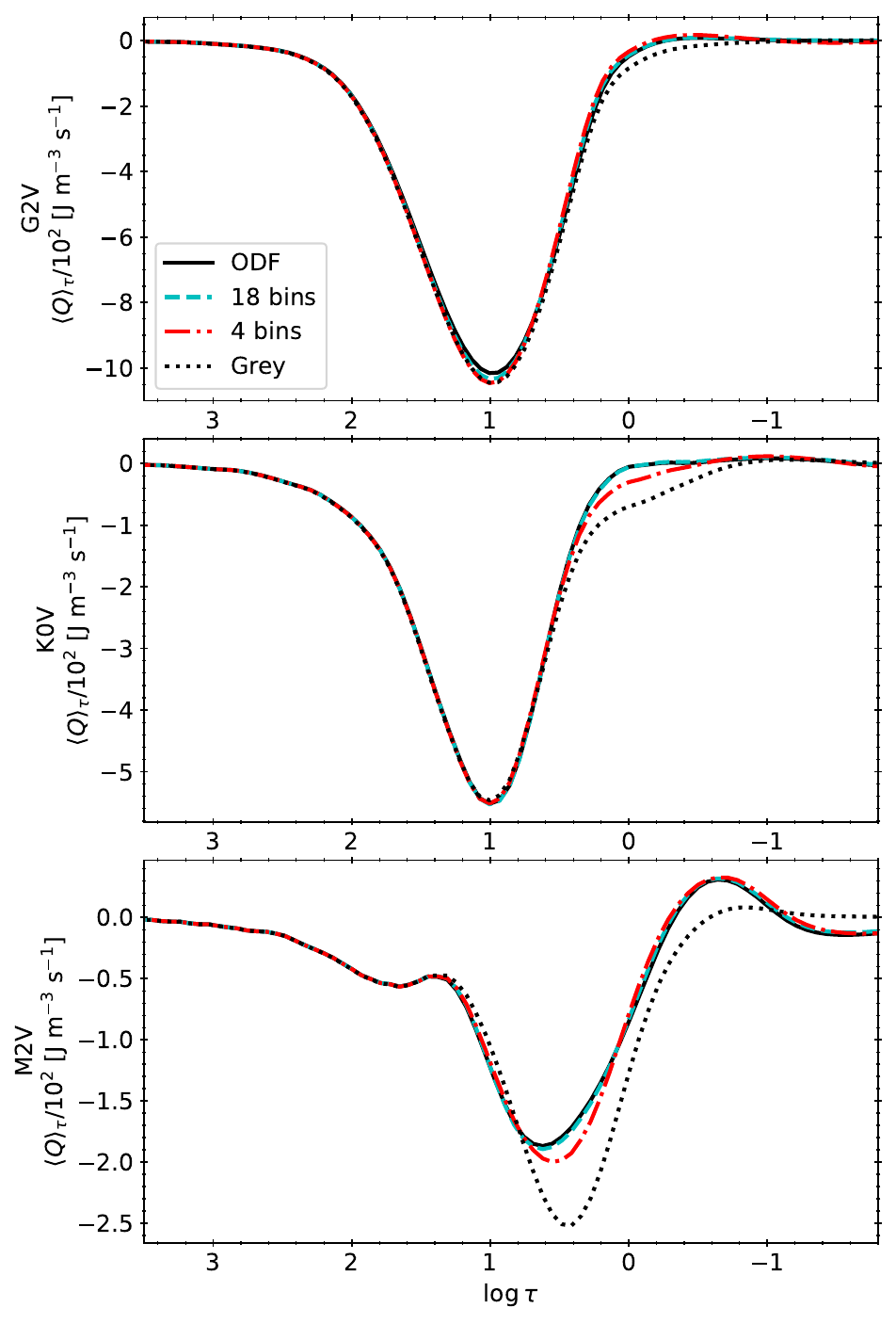}
        \caption{Radiative energy exchange rates $Q$ averaged over Rosseland iso-$\tau$ surfaces and computed using the ODF (solid black curve), the opacity labelled as \obd\ (dash blue curve), \obc\ (dash-dotted red curve), and \grey\ (dotted black curve), for the G2V (top column), K0V (middle), and M2V star (bottom).}
        \label{fig:Q1d_1snap}
    \end{figure}

    We select the last saved snapshot of the \grey\ runs of the three stars and 
    calculate $Q$ using the four opacity sets described in the App. \ref{app:bins}. 
    To compute $Q$, we apply the 3D RTE solver from MANCHA (see Sect.\ \ref{subsec:convection_mancha}) a posteriori in each of the selected snapshots. 
    The $Q$ rate calculated with the ODF setup is adopted as the reference instead of the exact monochromatic solution that is not computationally feasible in 3D. The ODF assumption is generally valid for the stellar atmospheres although it may break in the case of strong velocity fields causing significant shifts of the spectrum or in the case of the opacity of certain absorbers changing with height. Nevertheless, since we are interested in bolometric $Q$ and not in the detailed spectrum, the ODF assumption is valid. 

    The $Q$ rate averaged over the surfaces of constant Rosseland optical depth for the four opacity setups and the three snapshots is shown in Fig.\ \ref{fig:Q1d_1snap}. To evaluate the quality of the different opacity setups, we use the deviation measures $\chi_\mathrm{C}$ and $\chi_\mathrm{H}$ already introduced in \citetalias{2023paperI} as
    \begin{equation} \label{eq:error_c}
        \mathrm{\chi_{\mathrm{C}}} = \frac{ A \left( \left | Q^{\mathrm{ODF}}  -  Q^{\mathrm{OB}} \right |  \right)_{\mathrm{C}} }{A \left( \left | Q^{\mathrm{ODF}}  \right | \right)_{\mathrm{C}}},
    \end{equation} 
    \begin{equation} \label{eq:error_h}
        \mathrm{\chi_{\mathrm{H}}} = \frac{ A \left( \left | Q^{\mathrm{ODF}}  -  Q^{\mathrm{OB}} \right |  \right)_{\mathrm{H}} }{A \left( \left | Q^{\mathrm{ODF}}  \right | \right)_{\mathrm{H}}},
    \end{equation}
    where $Q^{\mathrm{ODF}}$ is the rate computed with the ODF and $Q^{\mathrm{OB}}$ is the rate computed with the multiple-bin or the grey opacity. The areas are defined as 
    \begin{equation}\label{eq:area_c}
        A \left( f(z) \right)_C = \int_{z_{\mathrm{b}}}^{z_{\mathrm{C \rightarrow H}}} f(z) dz, 
    \end{equation}
    \begin{equation}\label{eq:area_h}
        A \left( f(z) \right)_H = \int_{z_{\mathrm{C \rightarrow H}}}^{z_{\mathrm{t}}} f(z) dz, 
    \end{equation}
    where $z$ can be any height coordinate (e.g. the geometrical height or the optical depth), the height at $z_{\mathrm{C \rightarrow H}}$ is the height where $Q^{\mathrm{ODF}}$ changes its sign close to the surface, and $z_{\mathrm{b}}$ and $z_{\mathrm{t}}$ are, respectively, the highest point in the atmosphere under the surface ($z<0$ in the plots) and the lowest point above the surface ($z>0$) where $\left| Q^{\mathrm{ODF}} \right|< 2 \times 10^{-4} \left| \min( Q ) \right|$ (see figure 6 in \citetalias{2023paperI}). 

    \begin{table}[h]
        \caption{Deviation measures $\chi_\mathrm{C}$ and $\chi_\mathrm{H}$ in percentage computed as in \citetalias{2023paperI} for the mean stratifications of $Q$ shown in Fig.\ \ref{fig:Q1d_1snap}.}
        \begin{tabular}{ccccccc}
        \hline
        \hline
        \multicolumn{1}{l}{} & \multicolumn{2}{c}{\obd} & \multicolumn{2}{c}{\obc} & \multicolumn{2}{c}{\grey} \\ \hline
                             & $\chi_\mathrm{C} $& $\chi_\mathrm{H}$ &$\chi_\mathrm{C}$& $\chi_\mathrm{H}$ & $\chi_\mathrm{C}$& $\chi_\mathrm{H}$ \\ \hline
        G2V                  & 1& 26                     & 2& 57                    & 6 & 167                 \\ 
        K0V                  & 1& 23                     & 4& 48                    & 10& 105                 \\ 
        M2V                  & 2& 6                      & 6& 12                    & 25& 301                 \\ \hline
        \label{tab:chis}
        \end{tabular}
    \end{table}

    \begin{figure*}
        \centering
        \includegraphics[width=17.6cm]{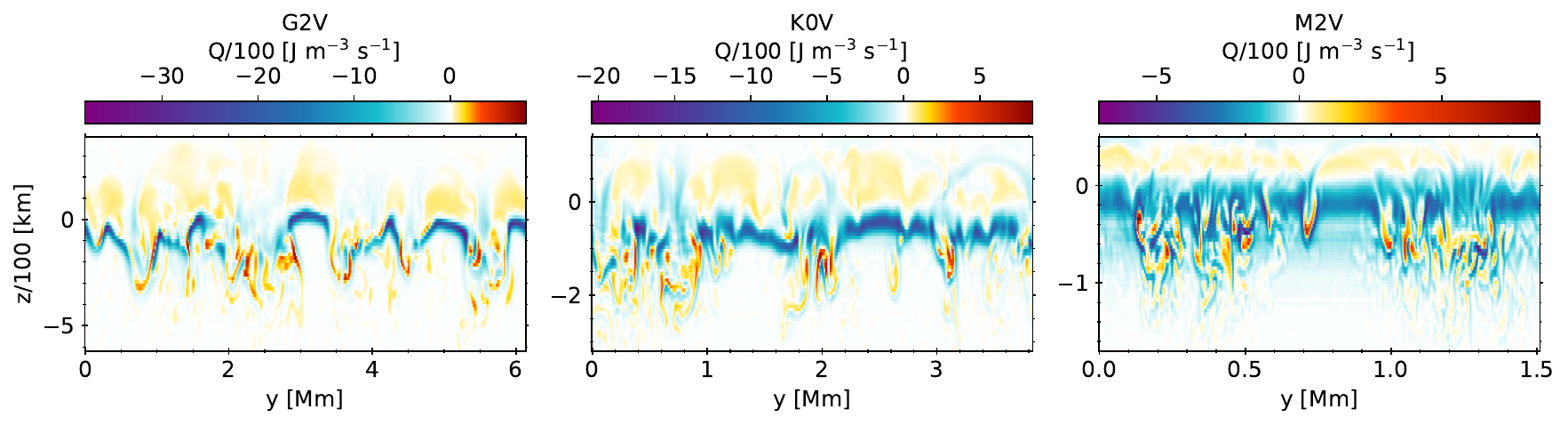}
        \caption{Colour maps of $Q$ computed with the ODF versus $y$ and $z$ in the same vertical cuts as in Fig.\ \ref{fig:stars_2d} for the last snapshot in the time series of the three stars (different panels). The corresponding colour-bar is shown over each panel. 
        }
        \label{fig:2DQ_ODF}
    \end{figure*}

    \begin{figure}
        \centering
        \includegraphics[width=7.cm]{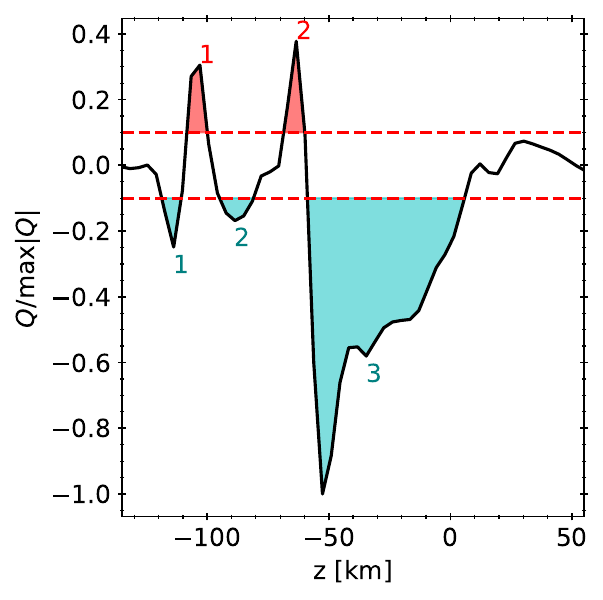}
        \caption{Example of identification of positive (red areas) and negative (blue areas) lobes in $Q/\max|Q|$ from the column at $x=0.88$ Mm and $y=0.73$ Mm for the M2V star. The horizontal dashed red lines mark the minimum amplitude required for a lobe to be considered in the count. }
        \label{fig:example_lobe_selection}
    \end{figure}

    The values of $\chi_\mathrm{C}$ and $\chi_\mathrm{H}$ for the nine cases in Fig.\ \ref{fig:Q1d_1snap} are listed in Table \ref{tab:chis}. For the three stars, the $Q$ from the \obd\ setup (dashed blue curve) and the one from ODF (solid black line) are nearly overlapping at all heights. The $Q$ rate computed with \obc\ (dashed-dotted red curve) is remarkably close to the previous two in the G2V case, showing only slightly more cooling at the minimum of the curve. The grey setup (dotted grey curve) for this star leads to $Q$ nearly identical to the one obtained with \obc\ at the curve minimum, but it shows weak cooling around $\log \tau=0$ instead of very weak heating predicted by the reference solution. 
    
    In the case of the K0V star the minimum of the curve is almost perfectly matched using the \obc\ setup. However, around $\log \tau =0$, $Q$ shows weak cooling while the reference solution in this case gives $Q$ virtually equal to zero. The grey solution replicates the same match at the minimum of the curve and around $\log \tau =0$ it gives the same trend as \obc\ but with even larger deviation. 

    \begin{figure*}
        \centering
        \includegraphics[width=14.cm]{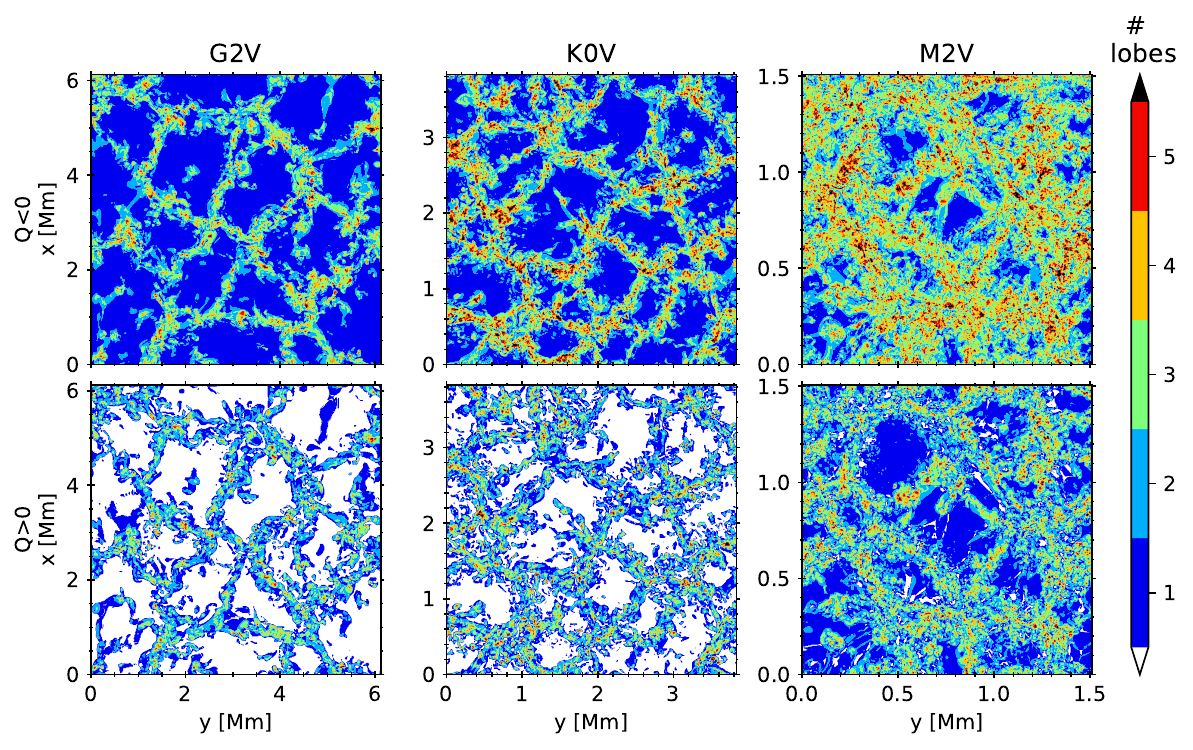}
        \caption{Colour maps of the number of lobes of $Q$ computed with the ODF in the columns of the last snapshot of the time series of the G2V (left column), K0V (middle), and M2V (right) star. The top row shows the number of minima with $Q<0$; the bottom row, the number of maxima with $Q>0$. Each colour corresponds to a different number of lobes, shown in the colour-bar at the right. The black colour in the colour-bar corresponds to more than five lobes, while white corresponds to zero.}
        \label{fig:number_of_lobes}
    \end{figure*}

    \begin{figure*}
        \centering
        \includegraphics[width=17.6cm]{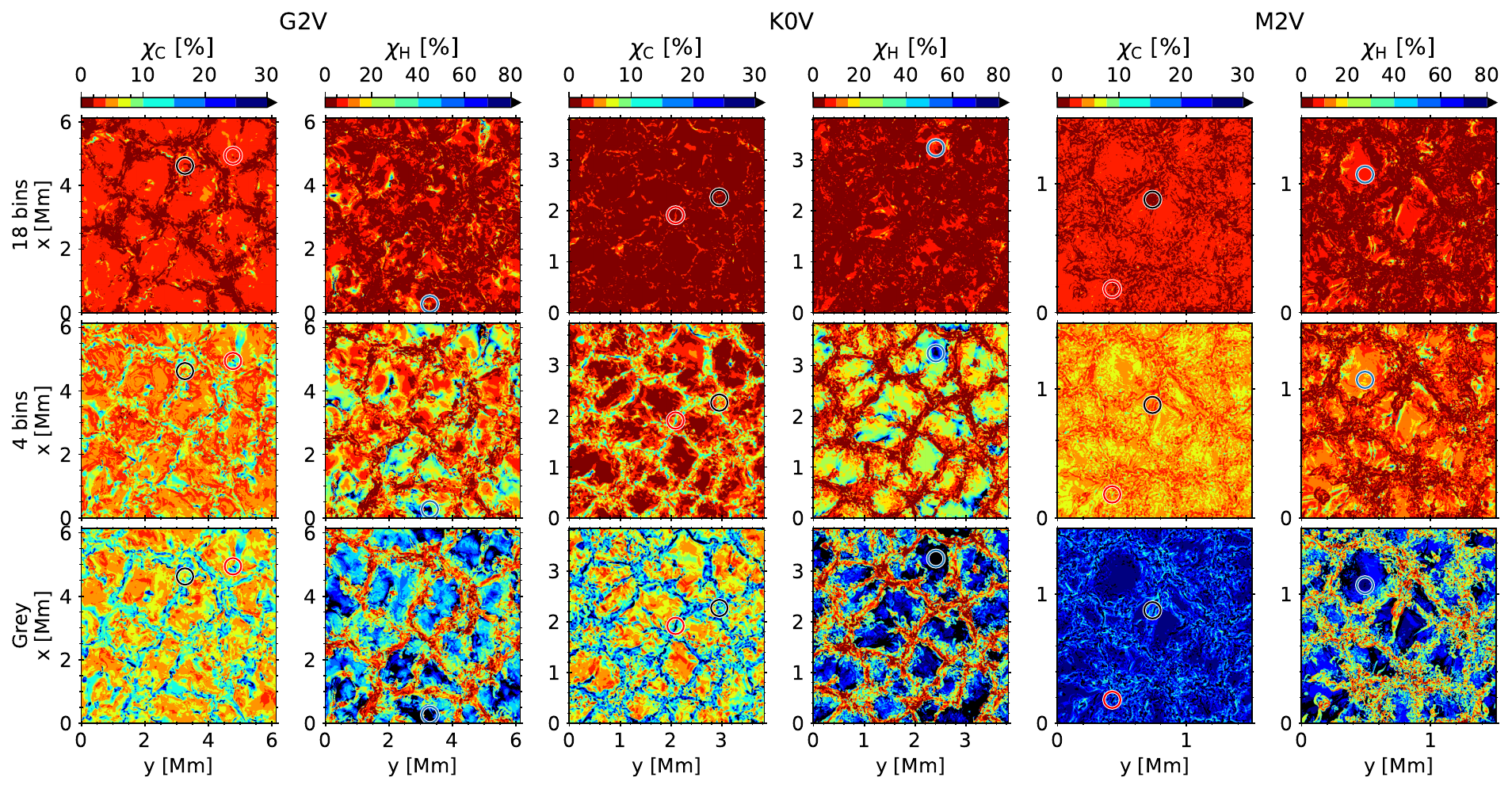}
        \caption{Colour maps of the deviation measures $\chi_\mathrm{C}$ (odd columns) and $\chi_\mathrm{H}$ (even columns) in the last snapshot of the time series of the G2V (left pair of columns), K0V (middle), and M2V (right) stars. The top row corresponds to the \obd\ opacity setup; the middle, to the \obc ; and the bottom, to the \grey . The corresponding discrete colour-bar is positioned over each column, ranging between $0$--$30\%$ and $0$--$80\%$ for $\chi_\mathrm{C}$ and $\chi_\mathrm{H}$, respectively. The black colour in the colour-bar corresponds to $\chi_\mathrm{C} > 30\%$ and $\chi_\mathrm{H} > 80\%$. 
        The division of the colour-scale is non-uniform.
        The circles mark the positions in the snapshots of the examples shown in Fig.\ \ref{fig:examples_of_error_1D} (the colour of the circles matches the colour of their corresponding curves in the figure).}
        \label{fig:distribution_of_chies}
    \end{figure*}

    \begin{figure*}
        \centering
        \includegraphics[width=17.6cm]{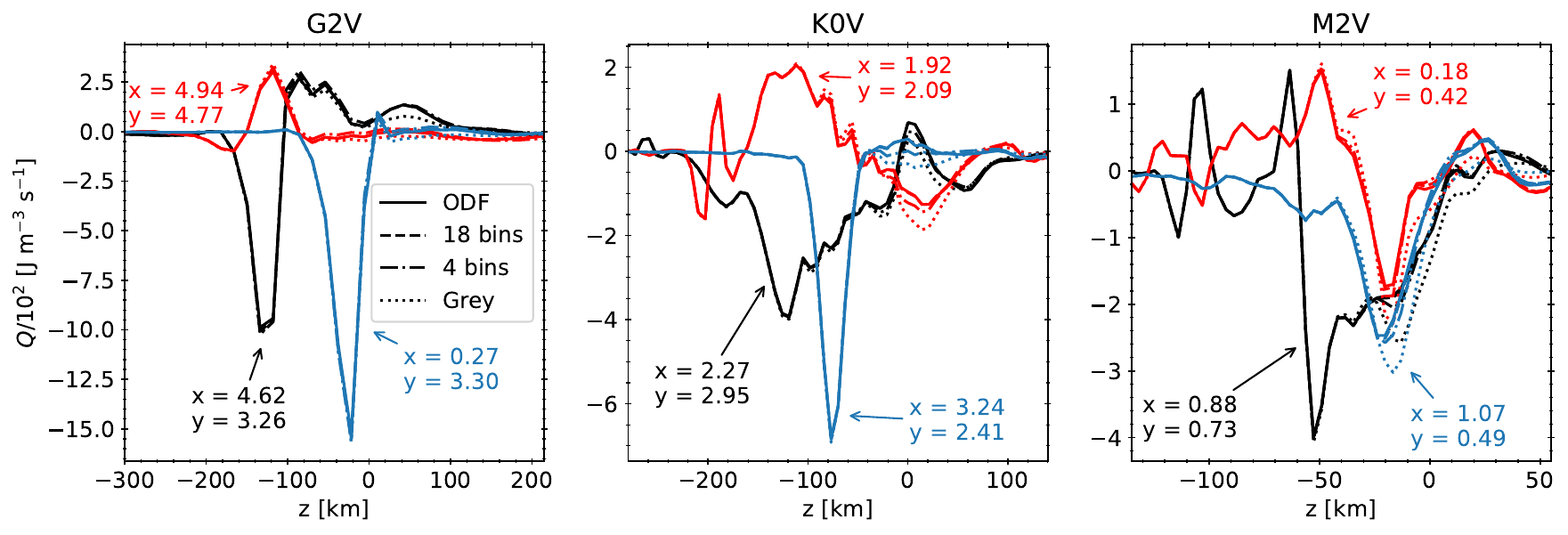}
        \caption{Radiative energy exchange rates in three columns of the last snapshot of the time series of the three stars (different panels). The red curves show an example of $Q$ in the intergranular lanes with large $\chi_\mathrm{C}$; the black curves show an example of $Q$ in the intergranular lanes with small $\chi_\mathrm{C}$; the blue curves show an example of $Q$ in the granules with large $\chi_\mathrm{H}$. Solid lines correspond to the ODF; dashed, to 18 bins; dashed-dotted, to four bins; and dotted, to grey opacity. The position $x$, $y$ of the selected columns in the stellar domain are written in Mm.}
        \label{fig:examples_of_error_1D}
    \end{figure*}
    
    In the plot for the M2V star 
    the \obc\ solution shows the opposite trend with respect to the one seen in the K0V star. The match to the reference $Q$ is perfect around $\log \tau=0$ but it shows significant deviation around the curve minimum. This stands in contrast to the conclusions of \citetalias{2023paperI} (see figure 11 there), where in the equivalent 1D case we found that the \obc\ solution gives considerably worse results. 
    For the M2V star, the \grey\ opacity setup leads to a large deviation of $Q$ at all heights, it overestimates the cooling component and it completely misses the heating one. This is due to underestimation of the line opacity, thus due to the reduction of the line blanketing with respect to the non-grey solutions. 

    The results for the three stars with only four bins show a significant improvement with respect to the results with the grey opacity, although they are not as good as those using 18 bins. 
    The choice of the binning strategy depends on the purpose and required accuracy of the simulation. 
    \citet{2004voegler_nongrey} concluded that it is sufficient to use the grey opacity for the study of the general properties of the solar granulation close to and under the surface. In the case of other stars, the applicability of the grey assumption for studying the general properties of stellar near-surface granulation depends on the relative importance of line blocking. For the M2V star studied in the present paper, the line blocking of molecules is so important that the cooling close to the surface would be highly overestimated if grey opacity was used (Fig.\ \ref{fig:Q1d_1snap}). Thus, depending on the spectral type, even for studying the near-surface convection, non-grey opacity might be needed. 
    Multiple opacity bins are required to produce simulations for detailed spectral synthesis, where accurate temperature gradients in the line forming regions are critical (see e.g. section 3.6.2 in \citealp{2012Freytag_co5bold}). In some cases, a few bins (usually four or five) are sufficient, for example in the study of line shifts and asymmetries of iron spectral lines 
    \citep{2000Asplund_ironI, 2000Asplund_ironII, 2001Shchukina_TrujilloB_iron}. 
    A higher number of bins ($\gtrsim 10$) is needed for the studies requiring high spectral sensitivity, such as solar CLV computations \citep{2023_muram_mps-atlas_12bins} or studies of the spatially resolved line profiles across stellar surfaces \citep{2016Dravins_exoplanets,2017Dravins_exoplanetsI}. Spectroscopic observations of the photospheric Fe{\tiny I} line profiles at different centre-to-limb positions were successfully obtained for real stars  \citep{2017Dravins_exoplanetsII,2018Dravins_exoplanetsIII}. These detailed observations provide a framework to test the accuracy of simulations and how well they describe real stellar surfaces \citep{2021Dravins_exoplanetsIV}. 
    
    Moreover, one should keep in mind that apart from the number of bins, the exact way how the opacities are distributed between the bins (i.e. the locations of the ${\tau,\lambda}$-separators) also matters. In particular, in
    \citetalias{2023paperI} it is shown that for the improvement of the RTE solution it is more important to carefully select the location of the bins in the $\tau,\lambda$-plane than to blindly increase the number of bins. A further improvement may be obtained following section 2.1.3 in \citet{beeck_thesis}. These authors proposed to employ the updated vertical stratification of the stellar snapshots to rebin opacity iteratively and they found that only one iteration is sufficient. 

    As shown at the top row of Fig.\ \ref{fig:up_down}, the final mean stratification of $Q$ in the G2V and K0V stars is mainly determined by the upflows. The $Q$ rate averaged only over the downflows in these snapshots is nearly zero. As the downflows cover only a small fraction of the surface, their contribution to the total mean of $Q$ is seen as a reduction of its amplitude. 
    In contrast to that, the mean $Q$ for upflows and downflows in the M2V star shows a similar trend, with upflows having a considerably higher amplitude. 
    This is because the difference in the temperature gradients of upflows and downflows (bottom row of Fig.\ \ref{fig:up_down}) is relatively smaller in the case of the M2V star with respect to the G2V and K0V star. This behaviour is consistent with their larger temperature contrast, as it is seen in Fig\ \ref{fig:tem_contrast}.

    Figure \ref{fig:2DQ_ODF} shows the colour-map for $Q$ computed with the ODF in the same vertical cuts as in Fig.\ref{fig:stars_2d}. As it has been already discussed in Sect. \ref{subsec:general_prop}, the granular tops contribute to the cooling in the final mean stratification, while the material above granules contribute to the heating due to the line blocking. This heating is better reproduced with ODF or with multiple bins than with grey opacity (e.g. compare Fig.\ \ref{fig:2DQ_ODF} with Figs.\ \ref{fig:stars_2d} and \ref{fig:compareQ_tau_npre}). 
    Figure \ref{fig:2DQ_ODF} also shows that there are small zones with strong heating concentrated in the intergranular lanes at relatively larger optical depths. Nevertheless, these heating zones are tiny compared to the large cooling areas, and therefore their contribution is small in the final mean stratification. The only heating that survives in the final mean stratification is that originating from above the granules.
    
    In 1D models $Q$ is calculated by solving RTE along a vertical ray. The resulting $Q$ stratification with height is usually simple showing one heating ($Q>0$) and one cooling ($Q<0$) lobe (figure 6 from \citetalias{2023paperI}). In 3D simulations $Q$ is obtained by solving RTE by ray tracing in space, and in every grid point of the 3D mesh the result is affected by the local inhomogeneities along the different rays. The consequence is that the $Q$ stratification per 1D column extracted from 3D snapshots shows complex profiles with height with multiple lobes. 
    To illustrate that complexity, for every column of the 3D snapshots we count the number of positive maxima and negative minima in the $Q$ (computed with the ODF) normalized by $\max|Q|$. The lobes with absolute amplitude lower than $0.1$ of the normalized $Q$ are discarded. 
    All the lobes that are surrounded by $|Q|/\max|Q|$ values with amplitude lower than $0.1$ are counted as a single lobe. An example of this lobe identification is shown in Fig.\ \ref{fig:example_lobe_selection}, where two heating and three cooling lobes are identified. In that example, the lobes between the two dashed red lines are not counted, and all the adjacent local minima in the range $z \in [-60,0]$ km (blue lobe 3) are counted as one lobe. 
    The number of lobes per column in the last snapshot in the time series of the three stars is shown in Fig.\ \ref{fig:number_of_lobes}. In all three stars the profiles in the granules resemble those from 1D models with one heating and one cooling lobe. In the case of the M2V star, the heating lobe is present in most of the granular columns, and the value of $Q/\max|Q|$ in these lobes surpasses the imposed threshold ($0.1$) which can be seen as dark blue regions mostly coinciding with granular tops (Fig.\ \ref{fig:example_lobe_selection}, the bottom-right panel). 
    In the G2V and K0V stars, if the heating lobe is present at all, it has amplitude $Q/\max|Q|<0.1$ (white regions in the bottom row of Fig.\ \ref{fig:example_lobe_selection}). 
 
    As explained above, the $Q$ rates in the columns of the 3D stellar domains in general do not present one cooling and one heating lobe, but show profiles with multiple lobes. Thus, to see which columns contribute more to the error in the final mean stratification, the deviation measures defined through the areas in Eqs.\ \ref{eq:area_c}-\ref{eq:area_h} are not useful. Instead, we generalize those definitions by changing the computation of the areas as
    \begin{equation}\label{eq:area_c_general}
        A \left( f(z) \right)_C = \sum_i \int_{\Delta z_i} f(z) dz, \; \mathrm{for} \; f(z \in \Delta z_i) < 0,
    \end{equation}
    \begin{equation}\label{eq:area_h_general}
        A \left( f(z) \right)_H = \sum_i \int_{\Delta z_i} f(z) dz,  \; \mathrm{for} \; f(z \in \Delta z_i) > 0,
    \end{equation}
    where $z$ can be any height coordinate (e.g. the geometrical height or the optical depth), the regions $\Delta z_i$ with positive or negative $Q$ are selected depending on the sign of $Q^{\mathrm{ODF}}$. 
    In the case of the $Q$ profiles with only one cooling and one heating lobe, these generalized deviation measures give the same values as those that use the areas from Eqs.\ \ref{eq:area_c},\ref{eq:area_h}.

    The distribution of $\chi_\mathrm{C}$ and $\chi_\mathrm{H}$ for the last snapshot of the time series of the three stars is shown in Fig.\ \ref{fig:distribution_of_chies} (first two columns correspond to G2V, second two to K0V, and third two to M2V star). 
    The deviations $\chi$ are clearly shaped by the granulation pattern. 
    For the \obd\ setup (top row in Fig.\ \ref{fig:distribution_of_chies}), the deviation $\chi_\mathrm{C}$ is larger in the granules than in the integranular lanes of the G2V and M2V snapshots. In the K0V star, the deviation $\chi_\mathrm{C}$ is larger in the integranular lanes than in the granules. In the three stars, the relatively large values of $\chi_\mathrm{H}$ are colocated with the granules, 
    above which the heating is significant (see e.g. Fig.\ \ref{fig:Q1d_1snap} and Fig.\ \ref{fig:2DQ_ODF}). We find $\chi_\mathrm{C}<5\%$ and $\chi_\mathrm{H}<10\%$ virtually everywhere in the maps of the three snapshots, hence the solution with 18 bins is very close to the ODF one. 

    For the \obc\ opacity setup (middle row in Fig.\ \ref{fig:distribution_of_chies}), 
    in the G2V and K0V stars the pixels with relatively large $\chi_\mathrm{C}$ are colocated with the intergranular lanes. Opposite to that, in the M2V star this correlation of $\chi_\mathrm{C}$ with the granulation is reversed and the deviation $\chi_\mathrm{C}$ is smaller in the intergranular lanes than in the granules. 
    In the three stars, the pixels with large $\chi_\mathrm{H}$ are again colocated with the granules and
    $\chi_\mathrm{C}<10\%$ virtually everywhere. For the M2V star, $\chi_\mathrm{H}<15\%$ in the majority of the snapshot, while in the G2V and K0V stars $\chi_\mathrm{H}<30\%$. Despite the larger values of $\chi_\mathrm{H}$ for lower number of bins, the heating in the G2V and K0V stars is negligible compared to the cooling. Although with less accuracy than the \obd\ setup, the \obc\ setup fairly reproduces the ODF solution. 

    Similarly to the case with \obc, the $\chi_\mathrm{C}$ for the \grey\ opacity setup (bottom row in Fig.\ \ref{fig:distribution_of_chies}) in the G2V and K0V stars is higher in the intergranular lanes than in the granules. Again, $\chi_\mathrm{C}$ in the M2V star shows a reversed correlation with the granulation with respect to the other two spectral types, with the granules having larger deviations than the intergranular lanes. As expected, the \grey\ opacity is the worst representation of the opacity as it produces very large deviations $\chi_\mathrm{H}$ everywhere in large granular areas. The areas of the granules have in most of the cases $\chi_\mathrm{C}<10\%$ in the case of the G2V and K0V. Since the granules are the regions that mainly determine the final mean stratification of $Q$ for these two stars, the grey approximation is still acceptable to study phenomena related to the convection in sub-- and near--surface layers which is consistent with the conclusions of \citealp{2004voegler_nongrey}. In the three stars, the deviation for the heating has values $\chi_\mathrm{H}>40\%$ almost everywhere in the map, implying that the grey opacity fails if the simulations are used for high precision spectroscopy.
    The large $\chi_\mathrm{H}$ is especially critical for the M2V star, for which the relative amplitude of the heating is comparable to that of the cooling. 
    Moreover, the pseudo-continuum in this star is strongly raised by the contributions from molecular lines into the opacity. These lines are underestimated in the grey opacity, thus the pseudo-continuum and line blanketing are underestimated too. The final result is that more cooling is produced with grey opacity than the one obtained with a binned opacity (cf. Fig.\ \ref{fig:Q1d_1snap} and Fig.\ref{fig:examples_of_error_1D}), 
    with $\chi_\mathrm{C}\in [20,30] \%$ in most of the snapshot. That makes the grey setup inappropiate for the M2V star, even for studies of near-surface convection. 

    To visualize better the description derived from Fig.\ \ref{fig:distribution_of_chies}, three examples of $Q$ in the columns of the 3D snapshots are shown in Figure \ref{fig:examples_of_error_1D}, computed with the four opacity setups (different line styles). Two of the selected columns are in the intergranular lanes with a relatively small and large $\chi_\mathrm{C}$ (black and red curves, respectively), and the third is in the middle of a granule, where $\chi_\mathrm{H}$ is larger (blue). Consistent with the description from Fig.\ \ref{fig:distribution_of_chies}, the largest errors in Fig.\ \ref{fig:examples_of_error_1D} found in the G2V and K0V stars (left and middle panels) correspond to the heating in atmospheric layers, while the cooling is mainly well represented by all the opacity setups. In the case of the M2V star (right panel in Fig.\ \ref{fig:examples_of_error_1D}), the multiple-bin opacities again produce close results to the results computed with the ODF, while $Q$ computed with grey opacities show larger errors for both the cooling and heating.


    \subsection{Effect of the binning on temperature stratification after temporal evolution} \label{subsec:Temperature}

    In this section, the simulations run with three opacity setups (\obd, \obc, and \grey) are analysed. 
    Regarding the results from Sect.\ \ref{subsec:Q}, it is clear that $Q$ computed with the \obd\ setup is closely overlapping $Q$ calculated with the ODF. Therefore, to address the impact of the different opacity setups on the mean stratification after time evolution, we adopt the simulations computed with \obd\ as the reference and compare them to those computed with \obc\ and \grey. 

\begin{figure*}
        \centering
        \includegraphics[width=15cm]{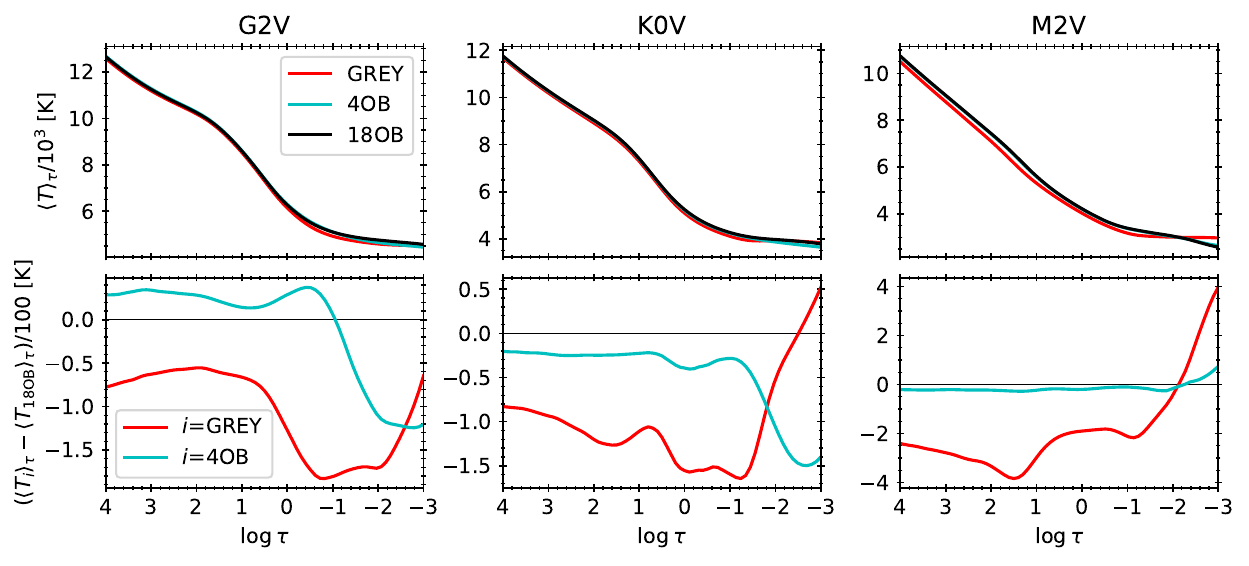}
        \caption{Mean stratification of temperature averaged over surfaces of the constant Rosseland optical depth and time of the G2V (left column), K0V (middle), and M2V (right) stars. 
        Top row: mean temperature $\langle T \rangle_\tau$ computed from the simulations run using the \obd\ (black curve), \obc\ (blue), and \grey\ (red) opacity setups. Bottom row: absolute difference of $\langle T \rangle_\tau$ computed with \obc\ (blue curve) and \grey\ (red) with respect to $\langle T \rangle_\tau$ obtained using \obd.}
        \label{fig:compareTemp}
    \end{figure*}

    The top row in Fig. \ref{fig:compareTemp} shows the mean stratification of the temperature $\langle T \rangle_\tau$ averaged over iso-$\tau$ surfaces and 
    time, for the three spectral types and three opacity setups. 
    In the bottom row, we show the difference between $\langle T \rangle_\tau$ computed with \obc\ and \grey\ opacity setups and $\langle T \rangle_\tau$ computed with \obd. 
    In the case of the G2V star, the mean atmosphere from the run computed with \obc\ is up to 40 K hotter than the one computed with \obd\ for $\log \tau>-0.5$. 
    Over this optical depth, this difference gradually decreases with height until
    $\langle T \rangle_\tau$ from the run with \obc\ becomes smaller than that computed with \obd\ (at $\log \tau <-1$). 
    The largest difference is for $\log \tau < -2$, where $\langle T \rangle_\tau$ from the run with \obc\ is around 120 K cooler than that with \obd. The resulting mean stratification from the run with \grey\ opacity has lower temperature than that with \obd\ for all heights in the atmosphere. For $\log \tau>0.5$ the difference between the stratifications ranges between 50 and 70 K. Higher up, the difference gradually increases up to 180 K at $\log \tau = -0.8$, it is maintained around 170 K until $\log \tau = -2$ and gradually decreases for $\log \tau < -2$, reaching around 60 K at $\log \tau = -3$.

    \begin{figure*}
        \centering
        \includegraphics[width=15cm]{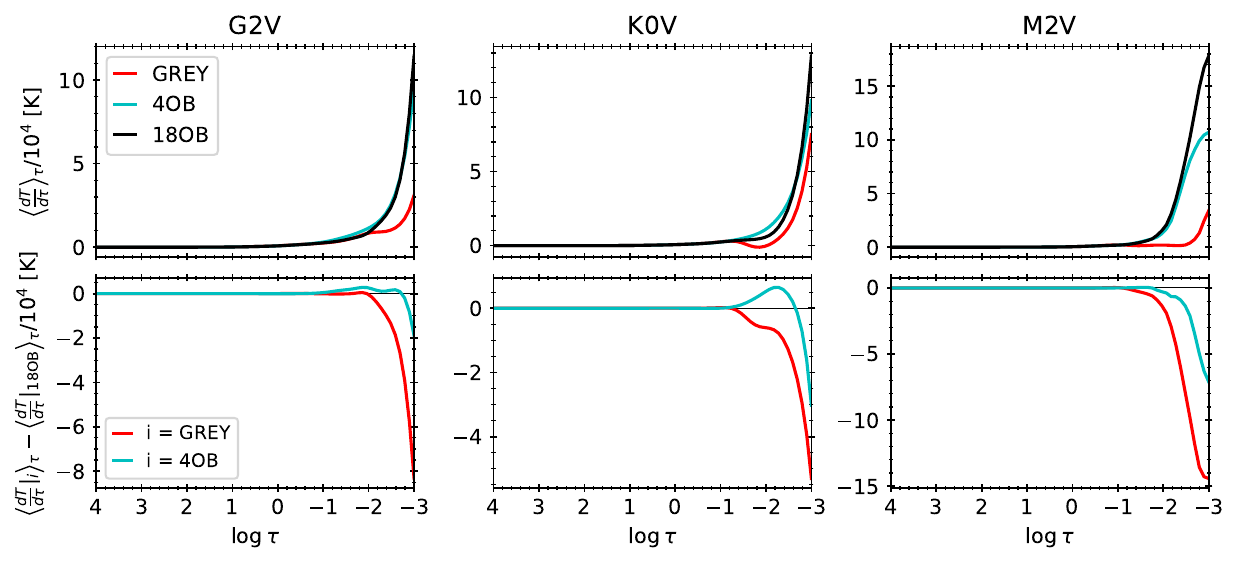}
        \caption{Mean stratification of the temperature gradient $\left\langle dT /d\tau \right\rangle_\tau$ averaged over surfaces of the constant Rosseland optical depth and time of the G2V (left column), K0V (middle), and M2V (right) stars. Top row: mean temperature gradient $\left\langle dT /d\tau \right\rangle_\tau$ computed from the simulations run using the \obd\ (black curve), \obc\ (blue), and \grey\ (red) opacity setups. Bottom row: absolute difference of $\left\langle dT /d\tau \right\rangle_\tau$ computed with \obc\ (blue curve) and \grey\ (red) compared to $\left\langle dT /d\tau \right\rangle_\tau$ obtained using \obd.}
        \label{fig:compareTempGrad_dtdtau}
    \end{figure*}
    
    \begin{figure*}
        \centering
        \includegraphics[width=15cm]{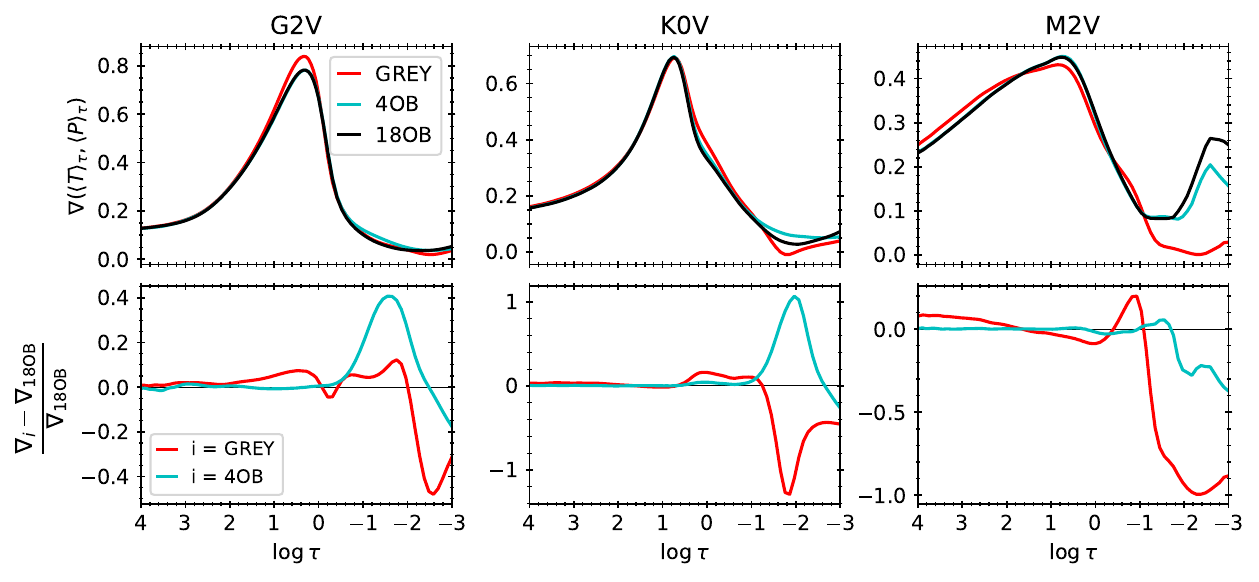}
        \caption{Temperature gradient averaged over time of the G2V (left column), K0V (middle), and M2V (right) stars. Top row: mean temperature gradient $\nabla = \partial \ln \langle T \rangle_\tau / \partial \ln \langle p \rangle_\tau$ computed from the mean stratification of temperature and pressure of the simulations run using the \obd\ (black curve), \obc\ (blue), and \grey\ (red) opacity setups. Bottom row: relative difference of $\nabla$ computed with \obc\ (blue curve) and \grey\ (red) compared to $\nabla$ obtained using \obd.}
        \label{fig:compareTempGrad}
    \end{figure*}

    The K0V star shows a mean stratification from the run with \obc\ with lower temperature from that with \obd\ at all heights. The difference is up to 40 K for $\log \tau >-1$, and it gradually increases up to around 150 K for $\log \tau<-2$. 
    The mean stratification from the run with \grey\ opacity is between 80 and 160 K cooler than that from the run with \obd\ opacity for $\log \tau > -1$. 
    Higher up in the atmosphere, the difference decreases and in $\log \tau \approx -2.5$ the temperature $\langle T \rangle_\tau$ from the \grey\ simulation becomes hotter than that from the \obd\ simulation, gradually increasing the difference with height up to 30 K at $\log \tau =-3$.

    For the M2V star, the mean stratification computed from the run with \obc\ and \grey\ are cooler than the one computed from the run with \obd\ for $\log \tau > -2$, with a difference up to 30 K for \obc\ and ranging between 200 K and 400 K for \grey. For $\log \tau < -2$, both of them become hotter than \obd, gradually increasing the difference up to 60 K and 390 K at $\log \tau=-3$ for \obc\ and \grey, respectively.

    Knowing the temperature gradient accurately is important for line synthesis and CLV computations \citep{book2003rutten}. The top row of Fig. \ref{fig:compareTempGrad_dtdtau} shows the mean stratification of the temperature gradient $\left\langle dT /d\tau \right\rangle_\tau$ averaged over iso-$\tau$ surfaces and time, for the three spectral types and three opacity setups. 
    The  bottom row of Fig. \ref{fig:compareTempGrad_dtdtau} shows the absolute difference $\left\langle \left. \frac{dT}{d\tau} \right|_i \right\rangle_\tau - \left\langle \left. \frac{dT}{d\tau} \right|_{\mathrm{18OB}}\right\rangle_\tau$ from the run with \obc\ and \grey\ with respect to that with \obd. 

    For the G2V star, the gradients computed from the \obc\ and \grey\ runs show the largest differences with respect to the gradient from the \obd\ run at $\log \tau < -1$. The difference at $\log \tau < -1$ is up to $-20000$ K for the \obc\ run, and up to $-85000$ K for the \grey\ run. 
    In the K0V star, both the \obc\ and \grey\ runs show significant differences with respect to the \obd\ run for $\log \tau < -1$, up to $-30000$ K and $-50000$ K, respectively.  
    In the M2V star, the gradient from the \grey\ run turns out significantly compromised with differences up to $-140000$ K at $\log \tau < -1$. This gap is largely reduced when the \obc\ setup is used, for which this difference is up to $-70000$ K at $\log \tau < -1$. It is thus evident that a non-grey RT is critical in the case of the M2V star. 

    The top row of Fig. \ref{fig:compareTempGrad} shows the temperature gradient $\nabla = \partial \ln \langle T \rangle_\tau / \partial \ln \langle p \rangle_\tau$ averaged over time, for the three spectral types and three opacity setups. The gradient is computed for the mean stratification of the temperature and pressure, averaged over iso-$\tau$ surfaces. It is computed this way to avoid the numerical ringing that would arise if the gradient was computed in the 3D snapshot column by column. 
    The bottom row of Fig. \ref{fig:compareTempGrad} shows the relative error $\left(  \nabla_i  - \nabla_\mathrm{18OB}  \right)/ \nabla_\mathrm{18OB}$ from the run with \obc\ and \grey\ compared to that with \obd. The variation of the relative error $\left(  \nabla_i  - \nabla_\mathrm{18OB}  \right)/ \nabla_\mathrm{18OB}$ with height is almost identical to the differences $\left\langle \left. \frac{dT}{d\tau} \right|_i \right\rangle_\tau - \left\langle \left. \frac{dT}{d\tau} \right|_{\mathrm{18OB}}\right\rangle_\tau$ described in \mbox{Fig.\ \ref{fig:compareTempGrad_dtdtau}.} 

    \begin{figure*}
        \centering
        \includegraphics[width=15cm]{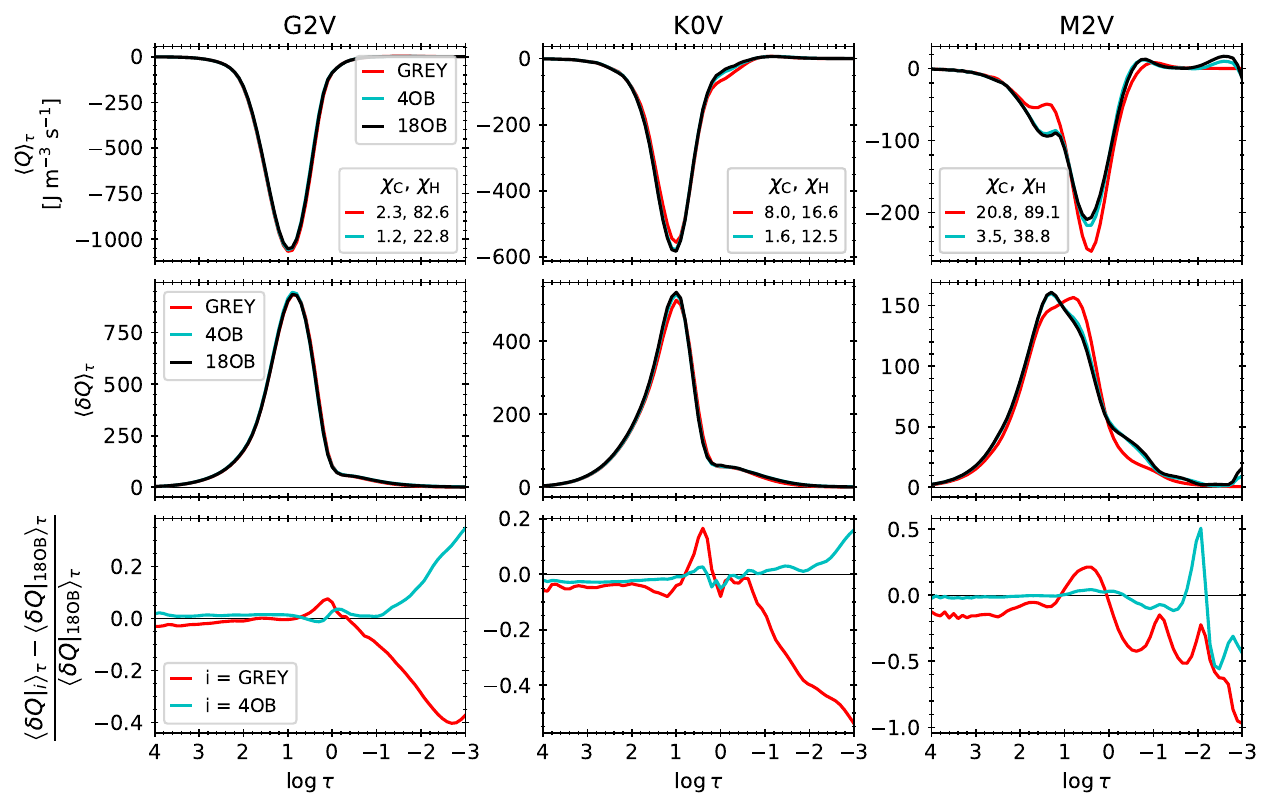}
        \caption{Mean stratification $\langle Q \rangle_\tau$ (top row) and fluctuations $\langle \delta Q \rangle_\tau$ (middle row) of the radiative energy exchange rate averaged over surfaces of the constant Rosseland optical depth and time of the G2V (left column), K0V (middle), and M2V (right) stars. The black, blue, and red curves in the top and middle row corresponds to the results computed from the simulations run using the \obd, \obc, and \grey\ opacity setups, respectively. The deviation measures $\chi_\mathrm{C}$ and $\chi_\mathrm{H}$ are shown in the legend at the panels of the top row. Bottom row: relative difference of $\left\langle \delta Q \right\rangle_\tau$ computed with \obc\ (blue curve), and \grey\ (red) compared to $\left\langle \delta Q \right\rangle_\tau$ obtained using \obd. The fluctuations shown in the middle row are computed as the standard deviation of $Q$.}
        \label{fig:Q_fluct}
    \end{figure*}

    \begin{figure*}
        \centering
        \includegraphics[width=15cm]{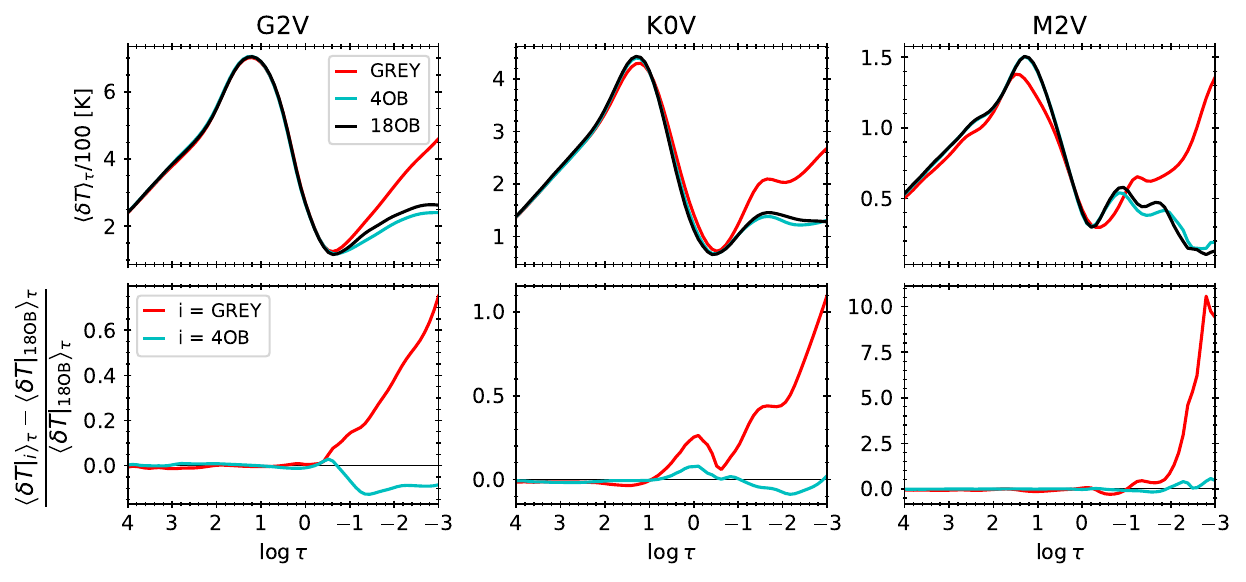}
        \caption{Mean stratification of the temperature fluctuations $\left\langle \delta T \right\rangle_\tau$ averaged over surfaces of the constant Rosseland optical depth and time of the G2V (left column), K0V (middle), and M2V (right) stars. Top row: mean temperature fluctuations computed from the simulations run using the \obd\ (black curve), \obc\ (blue), and \grey\ (red) opacity setups. Bottom row: relative difference of $\left\langle \delta T \right\rangle_\tau$ computed with \obc\ (blue curve) and \grey\ (red) compared to $\left\langle \delta T \right\rangle_\tau$ obtained using \obd. The fluctuations shown in the top row are computed as the standard deviation of the temperature.}
        \label{fig:temp_fluct}
    \end{figure*}

    The radiative energy exchange rate $\langle Q \rangle_\tau$ averaged over iso-$\tau$ surfaces and time is shown at the top row of Fig.\ \ref{fig:Q_fluct} for the three spectral types and three opacity setups. In each panel, the deviations $\chi$ (Eqs. \ref{eq:error_c},\ref{eq:error_h}) are shown in the legend. The deviations are computed using the $Q$ rate from the \obd\ runs as the reference (instead of $Q^{\mathrm{ODF}}$). As expected, for the M2V star the deviations are significantly reduced for the \obc\ case with respect to the \grey\ case, while for the G2V and K0V stars the difference is milder (particularly below the surface).

    The fluctuations of the temperature and $Q$ rate reflect how strong the coupling of the matter and radiation is in an atmosphere \citep[see e.g. chapter 5.3 in][for the case of the Sun]{voegler_thesis}. The temperature fluctuations $\langle \delta T \rangle_\tau$ averaged over iso-$\tau$ surfaces and time are shown in the top row of \mbox{Fig.\ \ref{fig:temp_fluct}} for the three spectral types and three opacity setups. The bottom row of Fig.\ \ref{fig:temp_fluct} shows the relative error $\left( \left\langle \left. \delta T \right|_i \right\rangle_\tau - \left\langle \left. \delta T \right|_{\mathrm{18OB}}\right\rangle_\tau \right) / \left\langle \left. \delta T \right|_{\mathrm{18OB}}\right\rangle_\tau$ from the run with \obc\ and \grey\ compared to that with \obd. The figure clearly shows for the three stars how the grey runs have larger temperature fluctuations than the \obc\ and \obd\ runs for $\log \tau >0$. This evinces a weaker coupling of the radiation field and matter for the grey runs, while in the \obc\ and \obd\ runs the radiative transfer decrease the amplitude of the fluctuations.

    The fluctuations of the $Q$ rate $\langle \delta Q \rangle_\tau$ averaged over iso-$\tau$ surfaces and time are shown in the middle panel of Fig.\ \ref{fig:Q_fluct} for the three spectral types and three opacity setups. 
    The bottom row of Fig.\ \ref{fig:Q_fluct} shows the relative error  
    $\left( \left\langle \left. \delta Q \right|_i \right\rangle_\tau - \left\langle \left. \delta Q \right|_{\mathrm{18OB}}\right\rangle_\tau \right) / \left\langle \left. \delta Q \right|_{\mathrm{18OB}}\right\rangle_\tau$ from the run with \obc\ and \grey\ compared to that with \obd. In this case, the analysis regarding the coupling of matter and radiation is not as evident as in the case of the temperature fluctuations, since the $Q$ fluctuations above the surface are very small. The amplitude of the relative error of the $Q$ fluctuations of the \obc\ runs with respect to the fluctuations of the \obd\ runs grows from less than $5\%$ in $\log \tau \simeq 0$ up to $35\%$, $15\%$, and $50\%$ at $\log \tau < -2$ for the G2V, K0V, and M2V stars, respectively. In the case of the \grey\ runs, there is a local increase of the relative errors in the optical depths where $\langle \delta Q \rangle_\tau$ peaks ($N_\mathrm{p} \in [0,1]$) and the relative errors below the surface are up to $5\%$, $5\%$, and $15\%$ for the G2V, K0V, and M2V stars, respectively. Above the surface, the amplitude of the relative errors for the \grey\ runs are up to $40\%$, $50\%$, and $100\%$ for the G2V, K0V, and M2V stars, respectively.

    \begin{figure*}
        \centering
        \includegraphics[width=16cm]{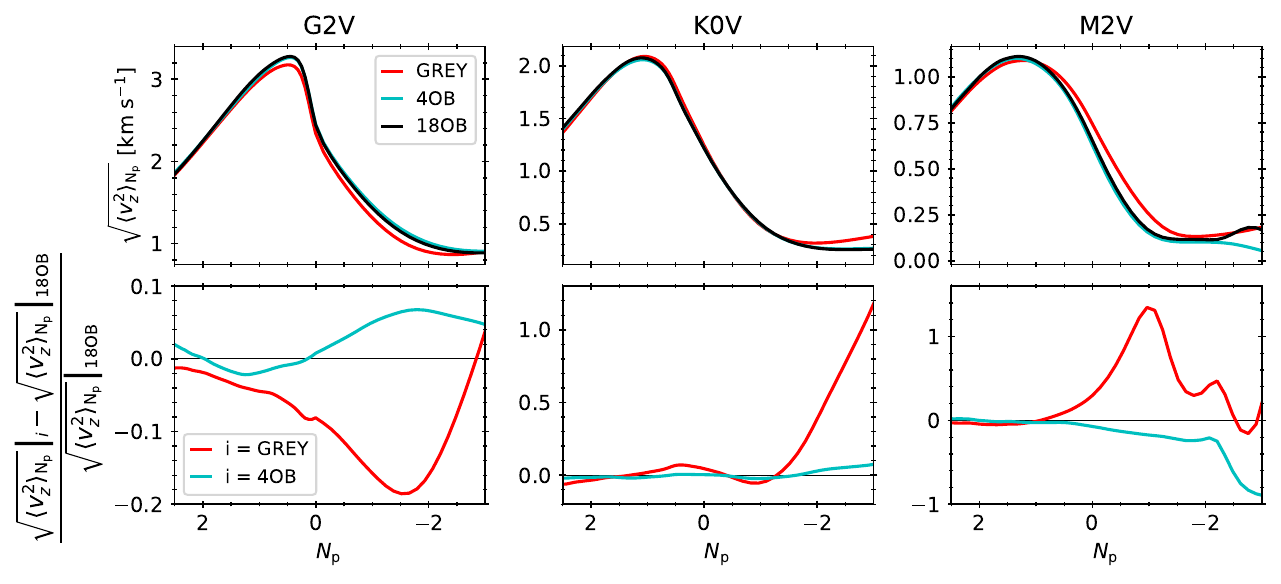}
        \caption{Mean stratification of RMS vertical velocity $\sqrt{ \left\langle v_z^2 \right\rangle_{\mathrm{N_p}} }$ averaged over surfaces of the constant $N_\mathrm{p}$ and time of the G2V (left column), K0V (middle), and M2V (right) stars. Top row: mean RMS vertical velocity computed from the simulations run using the \obd\ (black curve), \obc\ (blue), and \grey\ (red) opacity setups. Bottom row: relative difference of $\sqrt{ \left\langle v_z^2 \right\rangle_{\mathrm{N_p}} }$ computed with \obc\ (blue curve) and \grey\ (red) compared to $\sqrt{ \left\langle v_z^2 \right\rangle_{\mathrm{N_p}} }$ obtained using \obd.}
        \label{fig:veloc_binnings}
    \end{figure*}

    \begin{figure*}
        \centering
        \includegraphics[width=16cm]{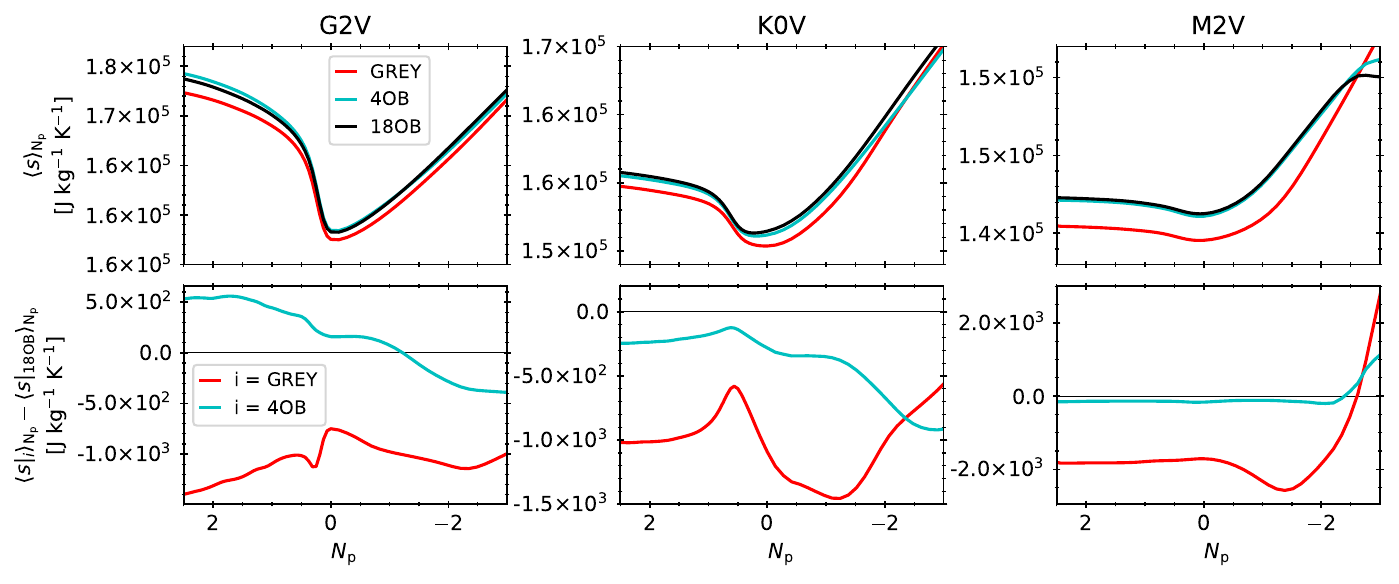}
        \caption{Mean stratification of the entropy $\left\langle  s \right\rangle_{\mathrm{N_p}} $ averaged over surfaces of the constant $N_\mathrm{p}$ and time of the G2V (left column), K0V (middle), and M2V (right) stars. Top row: mean entropy computed from the simulations run using the \obd\ (black curve), \obc\ (blue), and \grey\ (red) opacity setups. Bottom row: absolute difference of $\left\langle  s \right\rangle_{\mathrm{N_p}}$ computed with \obc\ (blue curve) and \grey\ (red) compared to $\left\langle  s \right\rangle_{\mathrm{N_p}}$ obtained using \obd.}
        \label{fig:entropy_binnings}
    \end{figure*}

    The RMS vertical velocity averaged over surfaces of the constant $N_\mathrm{p}$ is shown in the top row of Fig.\ \ref{fig:veloc_binnings} for the three spectral types and three opacity setups. The bottom row of the figure shows the relative error  $\left . \left( \left. \sqrt{ \left\langle v_z^2 \right\rangle_{\mathrm{N_p}} } \right|_i  - \left. \sqrt{ \left\langle v_z^2 \right\rangle_{\mathrm{N_p}} } \right|_{\mathrm{18OB}}  \right) \right/ \left. \sqrt{ \left\langle v_z^2 \right\rangle_{\mathrm{N_p}} } \right|_{\mathrm{18OB}} $ from the run with \obc\ and \grey\ compared to that with \obd. The RMS vertical velocity shows similar variation with height for the different opacity setups in the three stars. Under the surface, the amplitude of the relative error from both the run with \obc\ and \grey\ compared to that with \obd\ is lower than $10\%$ for the three stars. Above the surface, the amplitude of the relative error from the \obc\ run is still lower than $10\%$ for the G2V and K0V stars, and increases from $\lesssim 20\%$ at $N_\mathrm{p} \in [0,-2]$ up to $\simeq 100\%$ at $N_\mathrm{p} \simeq -3$. The relative error from the \grey\ run is up to $\simeq 20\%$ at $N_\mathrm{p} \simeq -1.5$ for the G2V star, up to $\simeq 100\%$ at $N_\mathrm{p} \simeq -3$ for the K0V star, and up to $\simeq 100\%$ at $N_\mathrm{p} \simeq -1$ for the M2V star. 
    
    Finally, in Fig.\ \ref{fig:entropy_binnings} we inspect the impact of the opacity setup on the entropy $\langle s \rangle_{\mathrm{N_p}}$ averaged over surfaces of the constant $N_\mathrm{p}$. 
    The entropy is computed from the EOS integrating the equation (see section 2.3.5 from \citealp{2002intro_sun_stix}):
    \begin{equation}
        ds = c_\mathrm{p} \left( \frac{dT}{T} -\nabla_{\mathrm{ad}} \frac{dp}{p} \right),
    \end{equation}
    where $c_\mathrm{p} = T \left( \partial s / \partial T \right)_\mathrm{p}$ is the heat capacity at constant pressure and $\nabla_\mathrm{ad} = \left( \partial \ln T / \partial \ln p \right)_\mathrm{p}$ is the adiabatic temperature gradient. The entropy is chosen to be zero at the lowest temperature and density  of the grid of the EOS table. Then, the entropy is interpolated in the domain of the G2V star and the value at the bottom of the domain is compared to that in figure 1a from \citet{1999Ludwig}. Finally, similarly to \citet{magic_thesis}, we add a bias to the entropy to cancel the difference between the entropy at the bottom of our G2V star and the entropy at the bottom of the Sun in \citet{1999Ludwig}, to have the same zero point.
    Figure\ \ref{fig:entropy_binnings} shows the entropy of the simulations run with each of the three opacity setups. 
    The bottom row of the figure shows the absolute difference $\left\langle \left. s \right|_i \right\rangle_{\mathrm{N_p}} - \left\langle \left. s \right|_{\mathrm{18OB}}\right\rangle_{\mathrm{N_p}}$ from the run with \obc\ and \grey\ compared to that with \obd. 
    There are two main effects on the entropy due to the changes in opacity. On the one hand, the entropy at the bottom of the domain is shifted when the opacity is changed, as it can be expected since the correction of the internal energy (or entropy) at the bottom boundary needs to be adapted for the changes in the radiative transfer that affect the stellar flux. The bias between the \grey\ and \obd\ cases is larger than that between the \obc\ and \obd\ cases for the three stars. On the other hand, the slope of the variation of the entropy with height may change for different opacity setups, as it is the case for the K0V and M2V simulations with \grey\ opacity, which shows a different slope for $N_\mathrm{p}<-1.5$ than that from the \obc\ and \obd\ runs. 
    
    To further quantify the change in the mean entropy between the runs with different opacity setups, Table\ \ref{tab:entropy_jump} shows the values of the so-called entropy jump, computed as the difference between the entropy at the bottom of the domain and the minimum of the entropy, 
    \begin{equation}
        \Delta s = \left\langle  s \right\rangle_{\mathrm{N_p}}|_{\mathrm{bottom}} - \left\langle  s \right\rangle_{\mathrm{N_p}}|_{\mathrm{min}} .
    \end{equation}
    In the three stars, the values of the entropy jump from the \obc\ runs are closer to the \obd\ case than the values from the \grey\ runs.
    
    The results from Figs. \ref{fig:compareTemp}-\ref{fig:entropy_binnings} are consistent with those in the previous section where we analysed a single snapshot per star. The \obc\ setup fairly represents the opacity in all three stars. The \grey\ setup leads to larger errors than the \obc\ one, especially for the M2V star. It should be noted that for this star the large difference in $\langle T \rangle_\tau$ below the surface for the \grey\ run with respect to the \obd\ one does not vary too much with height and it can be seen as a shift of the entire atmosphere in the Rosseland $\tau$ scale. Above the surface there is an increasingly large difference between temperature gradients obtained with the \grey\ setup with respect to the one from the \obd\ run. This difference is due to the underestimation of the line opacity in the grey setup. 

    \begin{table}[h]
        \caption{Entropy jump computed from the mean entropy shown in Fig.\ \ref{fig:entropy_binnings} for the three stars and three opacity setups.}
        \begin{tabular}{cccc}
        \hline
        \hline
              & \multicolumn{3}{c}{$\Delta s$ [J kg$^{-1}$ K$^{-1}$]} \\ \hline
              & \grey            & \obc              & \obd \\ \hline
        G2V   & $1.48\times10^4$ & $1.58\times10^4$  & $1.54\times10^4$   \\ 
        K0V   & $4.38\times10^3$ & $4.43\times10^3$  & $4.44\times10^3$   \\ 
        M2V   & $0.94\times10^2$ & $1.05\times10^3$  & $1.04\times10^3$   \\ \hline
        \label{tab:entropy_jump}
        \end{tabular}
    \end{table}

\section{Conclusions} \label{sec:conclusions}

    In this work, we present the first realistic 3D hydrodynamic simulations of near-surface convection in main sequence cool stars other than the Sun with the \texttt{MANCHA} code. We focus on three stellar types: G2V, K0V, and M2V. 
    Our results are consistent with the pioneering works of \citet{beeck_thesis} and \citet{magic_thesis}. 
    As a novelty with respect to the previous studies, the differences in the spatial distribution of the radiative energy exchange rate $Q$ between the three spectral types are studied in detail both in 1D mean stratifications and in 3D simulations. In particular, the relation between radiatively cooling and heating regions, and the granular structure is described. Our analysis of the uncertainties of different opacity binning setups for 3D simulations of the stellar granulation complements previous works and provides guidelines for efficient and accurate computing strategies.

    Four opacity setups are used to compute $Q$ from one snapshot of the time series of the three spectral types (Sect.\ \ref{subsec:Q}): ODF (as the reference), Rosseland opacity, and two opacities grouped in multiple bins (with four and 18 bins) that were confirmed to significantly reduce the error in the RTE solution in \citetalias{2023paperI}. In general, the $Q$ computed with 18 bins is remarkably close to that computed with the ODF, followed by the one calculated with four bins, which represents a significant improvement with respect to the grey solution. 
    In the G2V and K0V stars, the grey opacity reproduces well the cooling close to the surface. This opacity yields significantly high values for $\chi_\mathrm{H}$ in the three stars, owing to the underestimation of the line opacity. These high errors are especially critical for the M2V star, for which
    the relative amplitude of the heating with respect to the cooling is larger than in the other two. The underestimation of molecular line blanketing in the M2V also produces an excess of cooling close to its surface. 

    Although our preliminary study in 1D (\citetalias{2023paperI}) suggested that the use of a large number of bins ($> 10$) was required to fairly reproduce the $Q$ rate of the M2V star, the present study in 3D shows that the $\chi$ values produced with four bins are close to those with 18 bins (Figs.\ \ref{fig:Q1d_1snap} and \ref{fig:distribution_of_chies}). 

    We studied the mean stratification of the temperature 
    (Sect. \ref{subsec:Temperature}) computed from the simulations using grey, four-bins and 18-bins opacity setups 
    during time evolution.
    The 18-bins opacity setup is adopted as the reference solution instead of ODF following the results from Sect.\ \ref{subsec:Q}. The difference in sub-surface layers of the mean stratification of the temperature of the \obc\ and the \obd\ run is less than 40 K for the three stars. This difference is larger in the atmosphere, reaching up to hundreds of Kelvins for $\log \tau <-2$ in the G2V and K0V stars and around 50 K at $\log \tau=-3$ for the M2V star. In general, the temperature difference between the \grey\ and the \obd\ runs is in the range between 50 and 150 K for $\log \tau > 0$ in the G2V and K0V stars. In the case of the M2V star, this difference is strikingly larger for the same range of heights, between 200 and 400 K, owing to a shift of the atmosphere in the Rosseland optical depth. For $\log \tau < 0$, the difference is up to 400 K in the M2V star, and lower than 200 K in the other two stars. 

    The mean stratification of the temperature gradient is also computed for the same three opacity runs per star. In general, the differences $\langle \nabla_i \rangle_\tau -\langle \nabla_\mathrm{18OB} \rangle_\tau$ between the gradients computed from the \obd\ and \obc\ or \grey\ runs are large close to the surface, where the relative error $\left( \langle \nabla_i \rangle_\tau -\langle \nabla_\mathrm{18OB} \rangle_\tau \right)/\langle \nabla_\mathrm{18OB} \rangle_\tau$ is small. In contrast, the relative errors are the largest for $\log \tau < -1$, where the difference is small. This is especially critical in the case of the M2V star, for which the relative error is up to $100\%$ for $\log \tau < -1$.

    Depending on the relative importance of the line blanketing, the grey opacity might suffice to study the general properties of the convection (as before stressed for the solar case by \citealp{2004voegler_nongrey}), or it may not be enough (as in the case of the M2V studied in the present work). 
    A non-grey representation of the opacity with a small number of bins (around four or five) can be appropriate in the case of the synthesis of some spectral lines \citep{2000Asplund_ironI, 2000Asplund_ironII} and for the production of grids of stars with varying spectral type and metallicity. More bins are needed 
    for the studies requiring high spectral sensitivity, such as solar CLV computations \citep{2023_muram_mps-atlas_12bins} or studies of the spatially resolved line profiles across stellar surfaces \citep{2016Dravins_exoplanets,2017Dravins_exoplanetsI}. 
    
    In \citetalias{2023paperI} we explored the impact of the number of bins and their location on the accuracy of $Q$ and concluded that the distribution of the bins is critical to get an accurate solution. Increasing the number of bins has a limited effect, and, over a certain number, the improvement gets smaller and the computational costs to solve the RTE increase. Moreover, there is not convergence into the ODF solution for an arbitrarily large number of bins because 
    the binning always mixes opacities from several wavelength ranges with different variation with height in a single bin. 
    In the present paper we selected a fixed set of opacity configurations to test their performance in 3D simulations. The best number and distribution of the bins depends on the requirements of the problem. 
    In practise, it is worth to carefully optimize the smallest number of bins possible, rather than increasing their number.

\begin{acknowledgements}
      This work was supported by the European Research Council through the Consolidator Grant ERC--2017--CoG--771310--PI2FA and by Spanish Ministry of Science through the grant PID2021--127487NB--I00.
      We acknowledge support from the Agencia Estatal de Investigación (AEI) of the Ministerio de Ciencia, Innovación y Universidades (MCIU) and the European Social Fund (ESF) under grant with reference PRE2018--086567.
      APG and NV are thankful for the support 
      by the European Research Council through the Synergy Grant number 810218 ("The Whole Sun", ERC-2018-SyG). 
      NV acknowledges financial support from the ERC AdG SUBSTELLAR, GA 101054354. 
      We thank the referee for the detailed and careful reading of the manuscript and for asking the questions that lead to the improvement of Sect.\ \ref{subsec:Temperature} and Appendix \ref{app:thermal_relaxation}. 
      The authors thankfully acknowledge RES resources provided by Barcelona Supercomputing Center in MareNostrum to the activity AECT-2020-1-0021 and the resources provided by centro de Supercomputación y Bioinnovación de la Universidad de Málaga in PICASSO to the activities AECT-2022-1-0019 and AECT-2022-2-0029. We also acknowledge the use of LaPalma supercomputer from the Instituto de Astrofísica de Canarias. 
      This research has made use of NASA's Astrophysics Data System Bibliographic Services. 
\end{acknowledgements}

\bibliographystyle{aa} 
\bibliography{references} 

\begin{appendix} 

\section{Distribution of the bins} \label{app:bins} 
    \FloatBarrier 

\begin{figure}[h]
        \centering
        \includegraphics[width=7.cm]{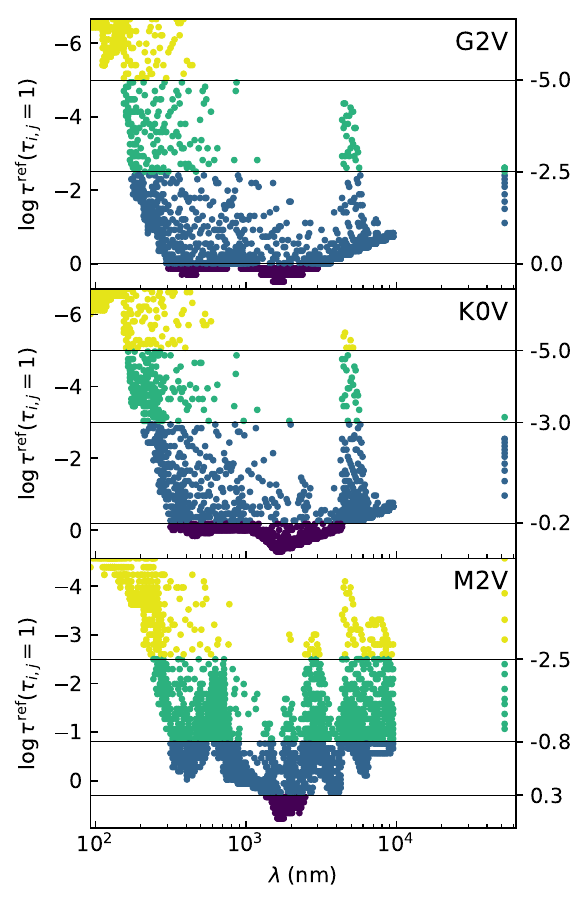}
        \caption{Reference optical depth and wavelength of the ODF points organized into four bins for the G2V, K0V, and M2V stars. As explained in \citetalias{2023paperI}, the Rosseland optical depth shown in the vertical axis is used as the reference $\tau^{\mathrm{ref}} (z_{i,j})$, where $z_{i,j}$ is the geometrical height at which $\tau_{i,j}=1$ for each step and substep of the ODF. The colours indicate different bins. Each point corresponds to one ODF substep, all centred in the middle wavelength of their corresponding step. The last column of dots at the right of each panel corresponds to the ODF step with $\lambda \in [9600, 9.5 \times 10^4]$ nm, and thus, appears separated from the rest of the dots.}
        \label{fig:four_bins}
    \end{figure}

\begin{figure}[h]
        \centering
        \includegraphics[width=7.cm]{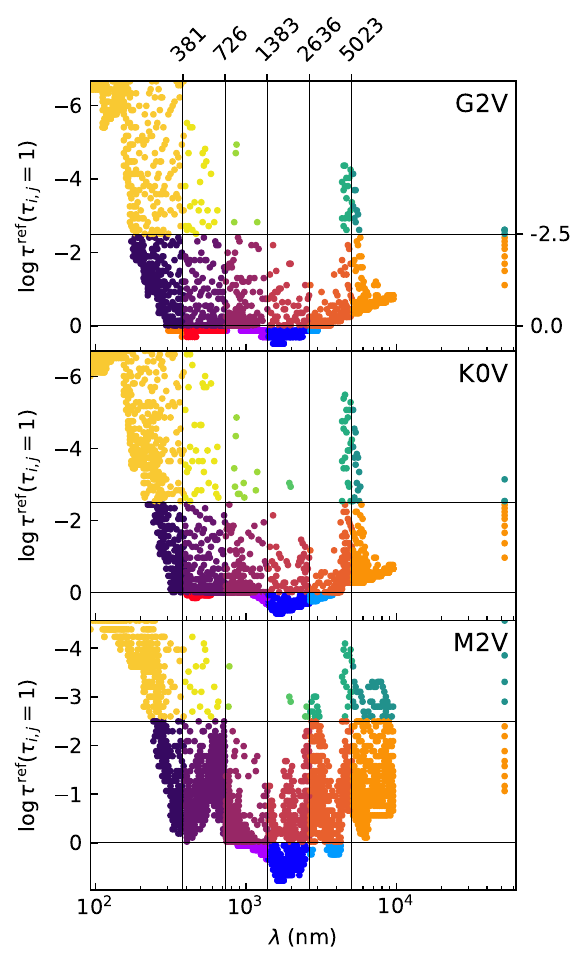}
        \caption{Same as Fig.\ \ref{fig:four_bins}, but for 18 bins. All the stars have the same $\{\tau, \lambda \}$ binning configuration.}
        \label{fig:18_bins}
    \end{figure}

    The four opacity setups used in this work were introduced in \citetalias{2023paperI}. The ODF is computed using the monochromatic opacities produced with \texttt{SYNSPEC}, for the same $T, \rho$ grids, wavelength steps and substeps described in the reference. 
    The binning for the selected four-bin and 18-bin opacity setups are shown in Figures \ref{fig:four_bins} and \ref{fig:18_bins}, respectively. These binning setups significantly reduce the errors of the RTE solution in 1D stellar models and, at the same time, have small enough number of bins so the computational cost is not too large in the case of 3D simulations. 
    The harmonic mean used to compute the grey Rosseland opacity is
    \begin{equation} \label{eq:harmonic_mean}
        \overline{\varkappa}_{\mathrm{Ro}} = \int \frac{\partial \mathrm{B}_\lambda}{\partial \mathrm{T}} d\lambda \Bigg / \left( \int  \frac{\partial \mathrm{B}_\lambda}{\partial \mathrm{T}} \frac{1}{\varkappa_\lambda} d\lambda \right),
    \end{equation}
    where $\partial \mathrm{B}_\lambda/\partial \mathrm{T}$ is the derivative of the Plank function, $\lambda$ is the wavelength, and $\varkappa_\lambda$ is the monochromatic opacity. This monochromatic opacity is the same one used to compute the ODF.

\clearpage
\section{Thermal relaxation of the simulations} \label{app:thermal_relaxation}

    \begin{figure*}[b!]
        \centering
        \includegraphics[width=16.cm]{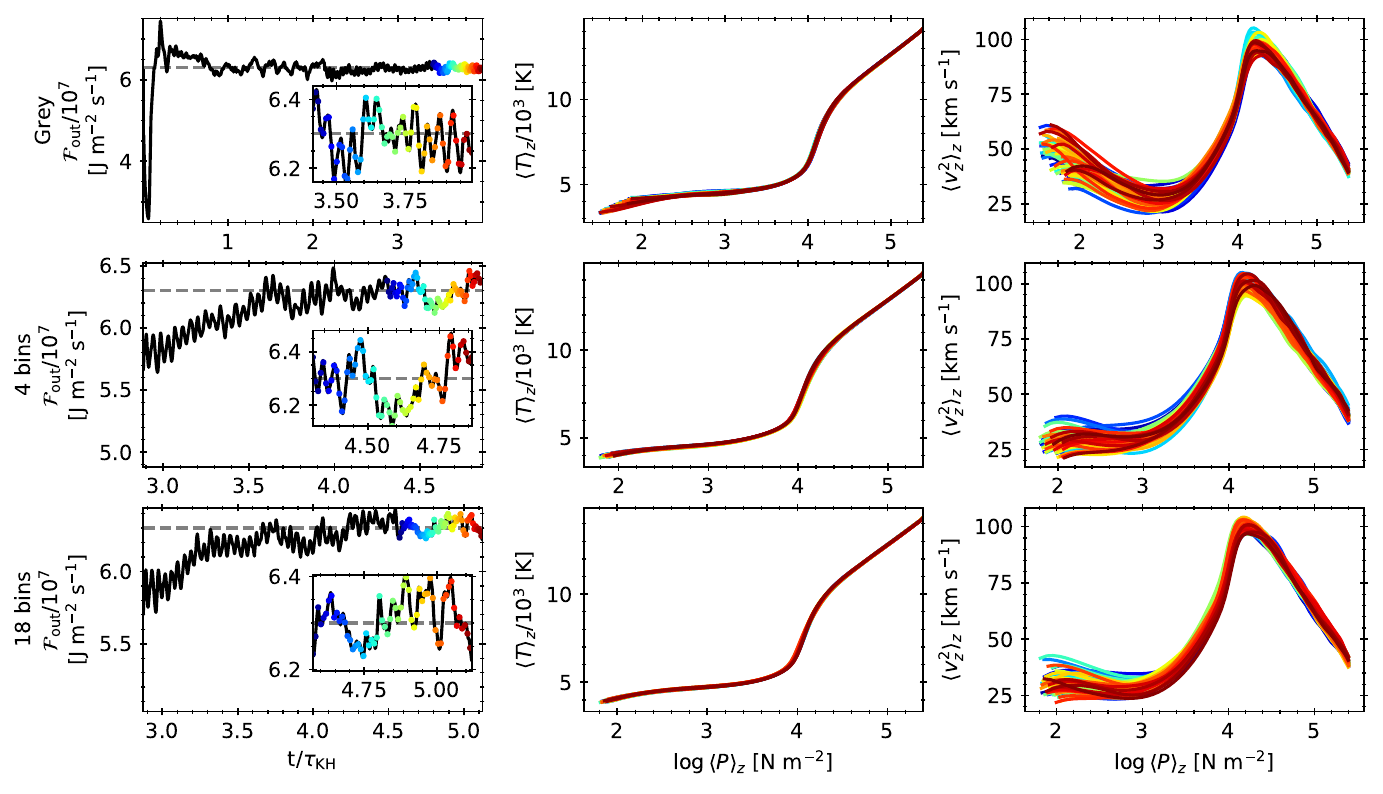}
        \caption{Emergent stellar flux (left column), mean stratification of temperature (middle), and RMS vertical velocity (right) averaged over planes of constant geometrical height for the simulations of the G2V star. Left column: the horizontal dashed grey line shows the targeted stellar flux (effective temperature, see Table \ref{tab:parameters_sim_1}). The horizontal axis shows the ratio of the time and the Kelvin-Helmholtz time scale. The coloured dots sample 50 snapshots of the time-series used in the results in the present paper. The corresponding embedded panels show the flux evolution for this time-series. Middle column: the temperature versus the logarithm pressure is shown at different times (with the same colours as the dots in the panels at the left column). Right column: the RMS vertical velocity versus the logarithm pressure is shown at different times (with the same colours as the dots in the panels at the left column). The top row corresponds to the simulation computed with \grey\ opacity; the middle row, with \obc; and the bottom row, with \obd.}
        \label{fig:thermal_g2v}
    \end{figure*}

\begin{figure*}[h]
        \centering
        \includegraphics[width=16.cm]{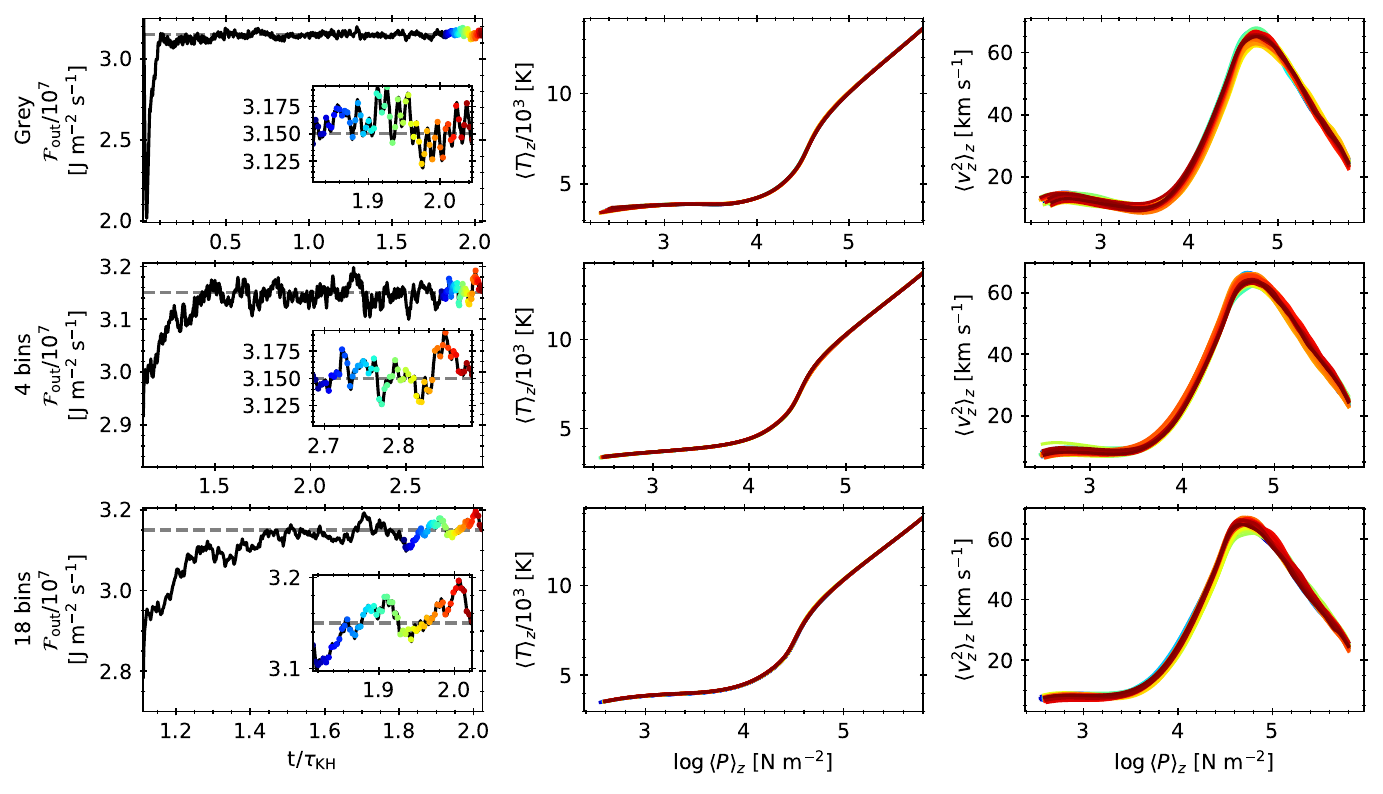}
        \caption{Same as Fig.\ \ref{fig:thermal_g2v}, but for the simulations of the K0V star.}
        \label{fig:thermal_k0v}
    \end{figure*}

\begin{figure*}[h]
        \centering
        \includegraphics[width=16.cm]{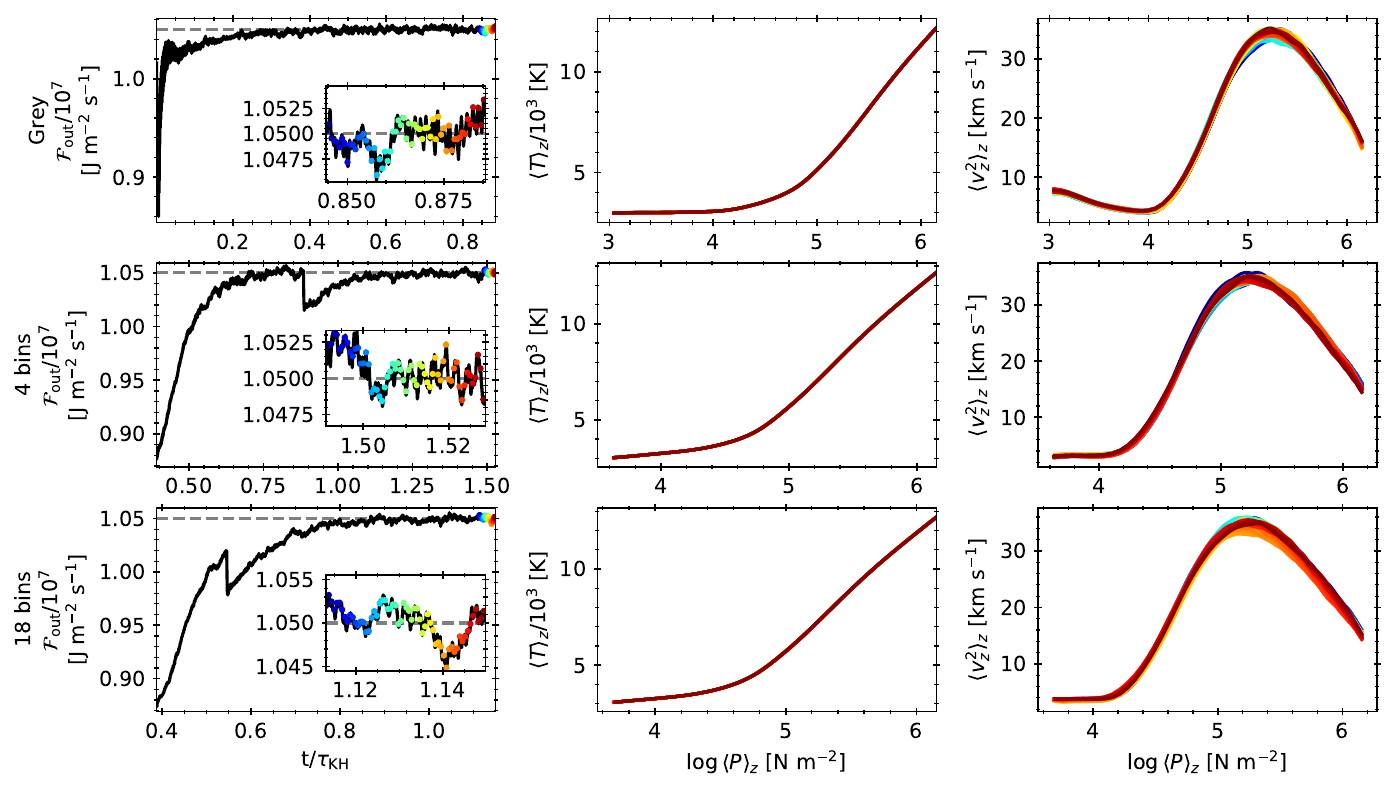}
        \caption{Same as Fig.\ \ref{fig:thermal_g2v}, but for the simulations of the M2V star. The sudden drop in the emergent flux of the middle and bottom panels was due to a change of opacity table. }
        \label{fig:thermal_m2v}
    \end{figure*}

    Figures \ref{fig:thermal_g2v}-\ref{fig:thermal_m2v} show that all the simulations used in this work are thermally relaxed. The left panels of the figures show the emergent stellar flux for the simulations run with \grey\ (top row), \obc\ (middle), and \obd\ (bottom) opacity setups. A larger time series than the one used for this paper is shown, needed to reach the stationary state. The x-axis shows the ratio of the time divided by the Kelvin-Helmholtz time scale, computed as (see equation 3.49 and related text in \citealp{voegler_thesis}):  
    \begin{equation}
        \tau_{\mathrm{KH}} = \frac{\int_{\mathrm{box}} e_{\mathrm{int}} dxdydz}{ \int_{\mathrm{top}} \mathcal{F} dxdy} ,
    \end{equation}
    where $e_{\mathrm{int}}$ is the internal energy per unit volume, $\mathcal{F}$ is the stellar flux, and $x,y,z$ are the three spatial coordinates. The subscript box means that the integral is done for the whole stellar domain, and the subscript top refers to an integral done in the top surface of the domain. 
    The middle panels of Figs.\ \ref{fig:thermal_g2v}-\ref{fig:thermal_m2v} show how the mean stratification of the temperature changes with time. Similarly, the right panels of the figures show how the RMS vertical velocity changes with time. 

\end{appendix}

\end{document}